\documentclass[sigplan,screen]{acmart}
\title{A Programming Model for Disaggregated Memory over CXL}
\usepackage{microtype}
\usepackage{graphicx} 
\usepackage{booktabs}
\usepackage{breqn}
\usepackage{amsthm}
\usepackage{pifont}
\usepackage{algorithm}
\usepackage{algpseudocode}
\usepackage[inline]{enumitem}
\algtext*{EndWhile}
\algtext*{EndIf}
\usepackage{listings}
\usepackage{parcolumns}
\usepackage{array}
\usepackage{multicol}
\usepackage{multirow}
\usepackage{xspace}
\usepackage{ifthen}
\usepackage{enumitem}
\usepackage[abbreviations]{foreign}  
\usepackage{xcolor}
\usepackage{colortbl}
\usepackage[skip=4pt,belowskip=0pt]{caption}
\usepackage{mathpartir}
\usepackage{subcaption}
\usepackage{mathtools}
\usepackage[capitalise,nameinlink]{cleveref}
\usepackage{marvosym}
\usepackage{tcolorbox}
\usepackage{subfiles}

\usepackage[export]{adjustbox} 
\usepackage{tabulary}       
\usepackage{makecell}       
\usepackage{tablefootnote}

\usepackage{eso-pic}

\AddToShipoutPictureBG*{%
  \AtPageUpperLeft{%
    \setlength\unitlength{1in}%
    \hspace*{\dimexpr0.5\paperwidth\relax}
    \makebox(0,-0.75)[c]{\textbf{Accepted to ASPLOS, 2026}}%
  }%
}

\Crefname{listing}{Algorithm}{Algorithms}
\crefname{listing}{Alg.}{Algs.}

\crefformat{section}{#2\S{}#1#3}
\Crefname{section}{Section}{Section}
\Crefformat{section}{Section #2#1#3}

\definecolor{codegreen}{rgb}{0,0.6,0}
\definecolor{codegray}{rgb}{0.5,0.5,0.5}
\definecolor{codepurple}{rgb}{0.58,0,0.82}
\definecolor{backcolour}{rgb}{0.95,0.95,0.92}

\makeatletter
\lst@AddToHook{OnEmptyLine}{\addtocounter{lstnumber}{-1}\vspace{-4pt}}
\makeatother

\author{Gal Assa}
\affiliation{
  \institution{Technion}            
  \country{Israel}                    
}
\email{galassa@technion.ac.il}

\author{Moritz Lumme}
\affiliation{
  \institution{ETH Zürich}            
  \country{Switzerland}                    
}
\email{moritz.lumme@inf.ethz.ch}

\author{Lucas Bürgi}
\affiliation{
  \institution{ETH Zürich}            
  \country{Switzerland}                    
}
\email{buergiluc@student.ethz.ch}

\author{Michal Friedman}
\affiliation{
  \institution{ETH Zürich}            
  \country{Switzerland}                    
}
\email{michal.friedman@inf.ethz.ch}

\author{Ori Lahav}
\affiliation{
  \institution{Tel Aviv University}            
  \country{Israel}                    
}
\email{orilahav@tau.ac.il}

\newcommand{\spec}{specification\xspace}

\newtheorem{definition}{Definition}

\newboolean{showcomments}
\setboolean{showcomments}{true}
\ifthenelse{\boolean{showcomments}}
{ \newcommand{\mynote}[3]{
				\fbox{\bfseries\sffamily\scriptsize#1}
				{\small$\blacktriangleright$\textsf{\emph{\color{#3}{#2}}}$\;\blacktriangleleft$}}}
{ \newcommand{\mynote}[3]{}}
\newcommand{\ga}[1]{\mynote{GA}{#1}{red}}
\newcommand{\mf}[1]{\mynote{Michal}{#1}{red}}
\newcommand{\ori}[1]{\mynote{ORI}{#1}{red}}
\newcommand{\ml}[1]{\mynote{Moritz}{#1}{red}}

\newcommand{\remove}[1]{}

\newcommand{\MODEL}{\textsf{CXL0}\xspace}

\newcommand{\conf}{\gamma}
\newcommand{\cache}{\mathit{C}}
\newcommand{\mem}{\mathit{M}}
\newcommand{\mycrash}[1]{\text{\Lightning}_{#1}}

\newcommand{\mycheck}{\textcolor{green!80!black}{\ding{52}}} 
\newcommand{\myxmark}{\textcolor{red}{\ding{55}}}

\newboolean{dontsavespace}
\setboolean{dontsavespace}{false}
\ifthenelse{\boolean{dontsavespace}}
{\newcommand{\mytext}[1]{#1}}
{\newcommand{\mytext}[1]{}}

\begin{document}
\begin{abstract}
    CXL (Compute Express Link) is an emerging open industry-standard interconnect between processing and memory devices that is expected to revolutionize the way systems are designed.
    It enables cache-coherent, shared memory pools in a disaggregated fashion at unprecedented scales, allowing algorithms to interact with various storage devices using simple loads and stores.
    While CXL unleashes unique opportunities, it also introduces challenges of data management and crash consistency. 
    For example, CXL currently lacks an adequate programming model, making it impossible to reason about the correctness and behavior of systems on top.

    In this work, we present \MODEL, the first programming model for concurrent programs over CXL.
    We propose a high-level abstraction for memory accesses and formally define operational semantics.
    We demonstrate that \MODEL captures a wide range of current and future CXL setups and perform initial measurements on real hardware.
    To illustrate the usefulness of \MODEL, we present a general transformation that enhances any linearizable concurrent algorithm with durability in a distributed partial-crash setting.
    We believe that this work will serve as a stepping stone for systems design and programming on top of CXL.
\end{abstract}




\maketitle
\section{Introduction}
With the increasing demand for storage in the cloud and recent improvements in interconnects~\cite{pond}, disaggregated memory has become a viable solution~\cite{cowbird,zhang2022optimizing,mind,apta,10.1145/3445814.3446713, lim2012implications, lerner2024vldb, ewais23survey, ding22hpc, maruf23challenges}.
In disaggregated systems, compute nodes are separated from memory nodes.
This offers greater flexibility, scalability, and resilience compared to traditional 
architectures~\cite{10.1145/3606557.3606563, ewais23survey}.
Rather than equipping a node to meet peak demand, it can be provisioned with enough memory to satisfy common needs, and additional memory is provided by other nodes during times of high load.
As a result,
data can be spread over multiple machines, and
data transfers can become a major bottleneck~\cite{lim2012implications, ewais23survey, dally2011power, dally2020accelerators}.

Compute Express Link (CXL) is an emerging standard which aims to facilitate efficient data transfer within data-centers.
Several cloud hyperscalers are involved in its development, and it is expected by many to become the future industry standard for data-center interconnects~\cite{cxl}.
CXL supports accessing external storage and caching data from remote nodes in a \emph{coherent} manner, which is a significant change compared to current architectures~\cite{CXLSpec40}. 
Furthermore, CXL provides a uniform abstraction for accessing all types of memory, such as DRAM and SSD, across the system via simple loads and stores at cache line granularity.
This allows the entire memory domain of the participating machines to be modeled as a large, unified shared main memory~\cite{lerner2024vldb}.

These opportunities also present new challenges.
As a centralized system becomes distributed \emph{and} coherent, two key assumptions no longer hold.
First, heterogeneous compute units access the same memory, so the outcome of concurrent operations is not governed by a single memory model.
Second, the system is no longer a single failure domain.
Instead, nodes may crash independently.
Special attention is required to ensure correctness under these conditions. 

Failure-aware programming has been subject to vast research effort, primarily in the context of full-system crashes~\cite{chandi1986dist, elnozahy2002dssurvey, koo1987checkdist, stevens16checkpoint}.
The emergence of non-volatile main memory (NVMM), as a persistent, moderately fast, and byte-addressable medium, gave rise to both theoretical and practical endeavors.
Correctness criteria \cite{izraelevitz2016linearizability,DBLP:conf/podc/AttiyaBH18,DBLP:conf/ppopp/FriedmanHMP18, li2021detectable, berryhill2016robust}, data structures \cite{DBLP:journals/pacmpl/ZurielFSCP19,
DBLP:conf/spaa/0001P21,DBLP:conf/ppopp/FriedmanHMP18,DBLP:conf/ppopp/Srivastava022}, 
verification approaches~\cite{AbdullaABKS21,DOsualdoRV23,BilaDLRW22,Kokologiannakis21,GorjiaraXD21,GorjiaraLLXD22}, and transactional memory frameworks \cite{ramalhete2021efficient,correia2018romulus,assa2023tl4x, 10.1145/3558481.3591079}
were suggested to benefit from the new technology. 
Although NVMM has been deprecated as a standalone product \cite{optane_nomore}, it is expected to remain part of the memory hierarchy with CXL~\cite{desnoyers2023persistent}. 
However, existing solutions are not automatically transferrable, because CXL inflates the system model, allowing multiple machines
to share the same volatile and persistent memory.
As these machines can be CPUs, GPUs, FPGAs, or any type of accelerator, an adequate programming model is required to correctly utilize this new technology.

Despite the expectation of CXL's wide adoption, there is currently no general programming model.
This leaves programmers
without answers to crucial questions.
For example, it is unclear what consistency guarantees memory accesses provide, or whether the local memory model is preserved over the communication protocol. 
Moreover, existing multi-threaded programs for shared memory do not remain correct in the disaggregated setting, and one urgent area of investigation is how to correctly port this legacy code.

This work aims to close these gaps.
We propose an abstraction of CXL's memory operations and define \MODEL as a formal programming model on top of that abstraction.
Both stem from a rigorous analysis of the CXL \spec~\cite{CXLSpec40}.
We capture a system model with distributed, coherent, shared memory, that allows for subsystem crashes.
Furthermore, we model each node in the system as a subsystem that emits sequences of CXL transactions, thus abstracting away the effect of the internal memory model.
Using \MODEL as a tool to reason about the correctness and behavior of concurrent executions, we show that it applies to a wide range of current and future CXL setups.
We also address the question of how to design concurrent programs for this new system model.
Shared memory programming in a multi-processor, disaggregated memory environment with independent failures is a challenging task.
We identify common pitfalls 
and demonstrate how to mitigate them using a general transformation that equips linearizable algorithms with fault tolerance under \MODEL.
Our transformation enhances legacy code with resilience to spontaneous subsystem failures, even for algorithms that are not failure-aware. 

\noindent
In summary, the contributions of this work are as follows:
\begin{itemize}[leftmargin=*,itemsep=0pt]
    \item We provide a high-level abstraction of memory accesses in the CXL specification;
    \item We introduce \MODEL, a programming model for coherent disaggregated memory over CXL;
    \item We present a transformation that makes linearizable code durable under the partial crash model and formally prove its correctness. 
\end{itemize}

To our knowledge, \MODEL is the first programming model for disaggregated memory over CXL, and our transformation is the first to equip linearizable code with durability in the partial crash model.
The rest of the paper is organized as follows.
We provide necessary background in \cref{sec:bg}.
In \cref{sec:model}, we describe our system model, abstraction and formalize the operational semantics that form \MODEL.
\cref{sec:newmodel} demonstrates how \MODEL captures current and future CXL setups, and \cref{sec:implications} illustrates practical implications on real hardware.
In \cref{sec:transformation}, we show how to transform existing concurrent programs for \MODEL.
\remove{ In \Cref{sec:proof}, we provide a correctness proof for these transformations. }
\cref{sec:related} reviews related work,
and \cref{sec:conc} concludes this paper.

The conference version of this paper is available in \cite{cxl0asplos}, and the code used for the proofs is available in \cite{cxl0supp}.
\section{Background}\label{sec:bg}

This section provides the essential background and preliminaries for this work.
We include a short introduction to the relevant parts of the CXL standard based on its \spec.

\paragraph{Disaggregated memory}
Disaggregated memory is a provisioning approach for improving hardware utilization in data centers.
The idea is to separate compute from memory, allowing each to scale independently.
It is an alternative to provisioning dedicated servers with enough resources to handle peak demand which leads to lower utilization during regular operation.
In addition to enabling independent scaling, disaggregated memory also allows independent maintenance 
and introduces a certain degree of fault tolerance (i.e., memory can outlast a compute unit accessing it).

\paragraph{Caching and cache coherence}
Memory caching techniques are used 
to increase efficiency by reducing latency when accessing memory.
This is accomplished by storing frequently used data in a small, high-speed memory close to the compute unit, which is much faster than main memory.
Modern processors are equipped with multiple layers of caches. 
The presence of caches means that the value of a logical memory address may reside in multiple physical locations (i.e., caches and main memory) simultaneously. 
Memory caches operate on fixed-size blocks called cache lines for data transfer.

Cache coherence is a system property that ensures \emph{each load operation observes the outcome of the most recent store to the same address that reached the cache}. 
Cache-coherent systems also satisfy \emph{single-writer multi-reader} exclusion and concurrent writes are impossible.
Many cache coherence protocols exist, most of which are aimed at single machines with multiple CPU cores and a hierarchy of shared and private caches.
Examples include MESI, MOESI, and MESIF, and some protocols for distributed settings \cite{apta,wang2021concordia}. 


\paragraph{Persistent memory programming}
Because systems rely on volatile caches, persistent memory introduces consistency challenges.
Stores are typically serviced by caches and are considered complete when they reach local cache rather than main memory.
When programming for volatile memory without considering crashes, cache replacement is usually irrelevant thanks to cache coherence.
However, when main memory is persistent, crash recovery depends on obtaining a consistent state from persistent memory.
Therefore, algorithms \emph{must} consider the order in which updates reach persistent media.
To that end, programmers use explicit \emph{flush} instructions, which propagate the content of a cache line to memory, and \emph{memory barriers} to enforce update ordering.
This ensures that the system can recover a consistent state from persistent memory regardless of when a crash occurs. 

\paragraph{Linearizability}
Consistency in the face of failures can manifest in various ways.
The canonical consistency criterion for concurrent objects and algorithms is \emph{linearizability}, as defined by \citet{herlihy1990linearizability}.
This criterion asserts that the concurrent history can be projected onto a single timeline in a manner that respects the sequential \spec.
The notion of linearizability has received multiple extensions in the context of partial and full-system crashes \cite{DBLP:conf/ppopp/FriedmanHMP18, aguilera2003strict,DBLP:conf/podc/AttiyaBH18, li2021detectable,guerraoui2004robust,berryhill2016robust}.
This work focuses on \emph{durable linearizability} by \citet{izraelevitz2016linearizability}.
Informally, the latter notion requires that completed operations persist before returning.
\subsection{A primer to CXL}

This work considers the most recent version of the CXL standard at the time of writing, CXL 4.0.
The \spec can be found in~\cite{CXLSpec40}.
The standard is as complex as one would expect, and a significantly shorter introduction to it was published by \citet{sharma2023introduction}.
Here, we describe only the details necessary for this work and refer the reader to either of the above sources for more information.

CXL consists of three sub-protocols: CXL.io, CXL.cache, and CXL.mem.
CXL.io, is a slightly augmented superset of PCIe that is used for
device discovery, configuration and power management.
CXL.cache implements a cache coherence domain, letting one PCIe endpoint (e.g., an accelerator) cache memory owned by a PCIe root complex (CPU).
CXL.mem provides the underlying memory channel for accessing remote memory via ordinary loads and stores.
The standard allows for different combinations of sub-protocols and defines three device types.
Type 1 devices run CXL.io alongside CXL.cache.
These devices only expose processing capabilities (e.g., SmartNICs).
Type 3 devices run CXL.io alongside CXL.mem.
These devices only expose device memory (e.g., main memory expanders).
This work focuses on Type 2 devices, which combine all three sub-protocols.
They expose both device memory and processing capabilities, such as GPUs, FPGAs and other accelerators.

All CXL sub-protocols are routable through custom CXL switches, so theoretically, the standard allows to combine dozens or even hundreds of root complexes (\emph{hosts}) and endpoints (\emph{devices}) into a single multi-root fabric in which memory devices can be pooled or shared.
In that setting, each host manages its own cache hierarchy and uses CXL.mem for access to remote memory.

Within a single-root coherence domain, e.g., the classic host-device pairing, CXL.cache implements MESI-based coherence on individual cache lines.
Importantly, it does not guarantee any ordering between accesses to memory locations on different cache lines.
The \spec suggests page-granular \emph{host-bias} and \emph{device-bias} modes for Type 2 device memory.
In host-bias mode, the host ``owns'' the memory page, meaning the device must request access from the host before accessing its own memory.
In device-bias mode, the device ``owns'' the page, and can access its memory directly without notifying the host.
These two modes represent a tradeoff between low-latency memory access for devices and cache-sharing overhead with the host, but they only apply within a single CXL.cache coherence domain. 

We note that the standard provides a variety of low-level transactions for reading, writing, and flushing cache lines, introducing a wide range of options for remote memory access that differ in performance and semantics.

\newcommand{\opsem}{
\begin{figure*}
\scalebox{0.95}{
\begin{mathpar}
\inferrule[\texttt{LSTORE}]{
\cache'_i = \cache_i[x \mapsto v] \\\\
\forall j \neq i \ldotp \cache'_j = \cache_j[x \mapsto \bot] 
}{\cache,\mem \xrightarrow{\texttt{LStore}_i(x,v)} \cache',\mem}
\and
\inferrule[\texttt{RSTORE}]{
x \in \mathit{Loc}_k \\\\
\cache'_k = \cache_k[x \mapsto v] \\\\
\forall j \neq k \ldotp \cache'_j = \cache_j[x \mapsto \bot] 
}{\cache,\mem \xrightarrow{\texttt{RStore}_i(x,v)} 
\cache',\mem}
\and
\inferrule[\texttt{MSTORE}]{
x \in \mathit{Loc}_k \\\\
\mem'_k = \mem_k[x \mapsto v] \\\\
\forall j \neq k \ldotp \mem'_j = \mem_j \\\\
\forall j \ldotp \cache'_j = \cache_j[x \mapsto \bot] 
}{\cache,\mem \xrightarrow{\texttt{MStore}_i(x,v)} \cache',\mem'}
\and
\inferrule[\texttt{Crash}]{
\cache'_i = \lambda x \ldotp \bot \\
\forall j \neq i \ldotp \cache'_j = \cache_j \\\\
\mem'_i = {\begin{cases}
\lambda x \ldotp 0 & i \text{ has volatile memory} \\
\mem_i & i \text{ has non-volatile memory}
\end{cases}} \\\\
\forall j \neq i \ldotp \mem'_j = \mem_j 
}{\cache,\mem \xrightarrow{\mycrash{i}} \cache',\mem'}
\and 
\inferrule[\texttt{LOAD-from-C}]{
\cache_j(x) = v \neq \bot \\\\
\cache'_i = \cache_i[x \mapsto v] \\\\
\forall j \neq i \ldotp \cache'_j = \cache_j 
}{\cache,\mem \xrightarrow{\texttt{Load}_i(x,v)} \cache',\mem}
\and
\inferrule[\texttt{LOAD-from-M}]{
\forall j \ldotp \cache_j(x) = \bot \\\\
x \in \mathit{Loc}_k \\ \mem_k(x) = v
}{\cache,\mem \xrightarrow{\texttt{Load}_i(x,v)} \cache,\mem}
\and
\inferrule[\texttt{Propagate-C-C}]{
i \neq k \\
x \in \mathit{Loc}_k \\
\cache_i(x) = v \neq \bot  \\\\
\cache'_i = \cache_i[x \mapsto \bot] \\\\
\cache'_k = \cache_k[x \mapsto v] \\\\
\forall j \not\in \{i,k\} \ldotp \cache'_j = \cache_j 
}{\cache,\mem \xrightarrow{\tau} \cache',\mem}
\and
\inferrule[\texttt{Propagate-C-M}]{
x \in \mathit{Loc}_k \\
\cache_k(x) = v \neq \bot \\\\
\forall j \ldotp \cache'_j = \cache_j[x \mapsto \bot] \\\\
\mem'_k = \mem_k[x \mapsto v] \\\\
\forall j \neq k \ldotp \mem'_j = \mem_j 
}{\cache,\mem \xrightarrow{\tau} \cache',\mem'}
\\
\inferrule[\texttt{LFLUSH}]{
\cache_i(x) = \bot
}{\cache,\mem \xrightarrow{\texttt{LFlush}_i(x)} \cache,\mem}
\and
\inferrule[\texttt{RFLUSH}]{
\forall j \ldotp \cache_j(x) = \bot
}{\cache,\mem \xrightarrow{\texttt{RFlush}_i(x)} \cache,\mem}
\and
\inferrule[\texttt{Global-Persistent-Flush}]{ 
\forall j,x \ldotp \cache_j(x) = \bot
}{\cache,\mem \xrightarrow{\texttt{GPF}_i} \cache,\mem}
%
\end{mathpar}
}
\caption{\MODEL: Operational Semantics (Load, Store, Crash, and Flush Primitives)}
\label{fig:opsem}
\end{figure*}
}

\newcommand{\newopsem}{
\begin{figure*}
\scalebox{0.9}{
\begin{mathpar}
\small
\raisebox{3.05em}{
    \inferrule[\texttt{LSTORE}]{
    \cache'_i = \cache_i[x \mapsto v] \\\\
    \forall j \neq i \ldotp \cache'_j = \cache_j[x \mapsto \bot] 
    }{\cache,\mem \xrightarrow{\texttt{LStore}_i(x,v)} \cache',\mem}
}
\and
\raisebox{1.95em}{
    \inferrule[\texttt{RSTORE}]{
    x \in \mathit{Loc}_k \\\\
    \cache'_k = \cache_k[x \mapsto v] \\\\
    \forall j \neq k \ldotp \cache'_j = \cache_j[x \mapsto \bot] 
    }{\cache,\mem \xrightarrow{\texttt{RStore}_i(x,v)} 
    \cache',\mem}
}
\and
\raisebox{0.6em}{
    \inferrule[\texttt{MSTORE}]{
    x \in \mathit{Loc}_k \\\\
    \mem'_k = \mem_k[x \mapsto v] \\\\
    \forall j \neq k \ldotp \mem'_j = \mem_j \\\\
    \forall j \ldotp \cache'_j = \cache_j[x \mapsto \bot] 
    }{\cache,\mem \xrightarrow{\texttt{MStore}_i(x,v)} \cache',\mem'}
}
\and
\inferrule[\texttt{Crash}]{
\cache'_i = \lambda x \ldotp \bot \\
\forall j \neq i \ldotp \cache'_j = \cache_j \\\\
\mem'_i = {\begin{cases}
\lambda x \ldotp 0 & i \text{ has volatile memory} \\
\mem_i & i \text{ has non-volatile memory}
\end{cases}} \\\\
\forall j \neq i \ldotp \mem'_j = \mem_j 
}{\cache,\mem \xrightarrow{\mycrash{i}} \cache',\mem'}
\\ 
\raisebox{1.6em}{
    \inferrule[\texttt{LOAD-from-C}]{
    \cache_j(x) = v \neq \bot \\\\
    \cache'_i = \cache_i[x \mapsto v] \\\\
    \forall j \neq i \ldotp \cache'_j = \cache_j 
    }{\cache,\mem \xrightarrow{\texttt{Load}_i(x,v)} \cache',\mem}
}
\and
\raisebox{2.85em}{
    \inferrule[\texttt{LOAD-from-M}]{
    \forall j \ldotp \cache_j(x) = \bot \\\\
    x \in \mathit{Loc}_k \\ \mem_k(x) = v
    }{\cache,\mem \xrightarrow{\texttt{Load}_i(x,v)} \cache,\mem}
}
\and
\inferrule[\texttt{Propagate-C-C}]{
i \neq k \\
x \in \mathit{Loc}_k \\
\cache_i(x) = v \neq \bot  \\\\
\cache'_i = \cache_i[x \mapsto \bot] \\\\
\cache'_k = \cache_k[x \mapsto v] \\\\
\forall j \not\in \{i,k\} \ldotp \cache'_j = \cache_j 
}{\cache,\mem \xrightarrow{\tau} \cache',\mem}
\and
\inferrule[\texttt{Propagate-C-M}]{
x \in \mathit{Loc}_k \\
\cache_k(x) = v \neq \bot \\\\
\forall j \ldotp \cache'_j = \cache_j[x \mapsto \bot] \\\\
\mem'_k = \mem_k[x \mapsto v] \\\\
\forall j \neq k \ldotp \mem'_j = \mem_j 
}{\cache,\mem \xrightarrow{\tau} \cache',\mem'}
\\
\inferrule[\texttt{LFLUSH}]{
\cache_i(x) = \bot
}{\cache,\mem \xrightarrow{\texttt{LFlush}_i(x)} \cache,\mem}
\and
\inferrule[\texttt{RFLUSH}]{
\forall j \ldotp \cache_j(x) = \bot
}{\cache,\mem \xrightarrow{\texttt{RFlush}_i(x)} \cache,\mem}
\and
\inferrule[\texttt{Global-Persistent-Flush}]{ 
\forall j,x \ldotp \cache_j(x) = \bot
}{\cache,\mem \xrightarrow{\texttt{GPF}_i} \cache,\mem}
\end{mathpar}
}
\caption{\MODEL: Operational Semantics (Load, Store, Crash, and Flush Primitives)}
\label{fig:opsem}
\end{figure*}

}

\newcommand{\adjopsem}{
\begin{figure*}
\centering
\scalebox{0.85}{
    \begin{minipage}{\linewidth}
    \centering
    \small  
    
    \begin{mathpar}
        \adjustbox{valign=t}{$
            \inferrule[\texttt{LSTORE}]{
            \cache'_i = \cache_i[x \mapsto v] \\\\
            \forall j \neq i \ldotp \cache'_j = \cache_j[x \mapsto \bot] 
            }{\cache,\mem \xrightarrow{\texttt{LStore}_i(x,v)} \cache',\mem}
        $}
        \and
        \adjustbox{valign=t}{$
            \inferrule[\texttt{RSTORE}]{
            x \in \mathit{Loc}_k \\\\
            \cache'_k = \cache_k[x \mapsto v] \\\\
            \forall j \neq k \ldotp \cache'_j = \cache_j[x \mapsto \bot] 
            }{\cache,\mem \xrightarrow{\texttt{RStore}_i(x,v)} 
            \cache',\mem}
        $}
        \and
        \adjustbox{valign=t}{$
            \inferrule[\texttt{MSTORE}]{
            x \in \mathit{Loc}_k \\\\
            \mem'_k = \mem_k[x \mapsto v] \\\\
            \forall j \neq k \ldotp \mem'_j = \mem_j \\\\
            \forall j \ldotp \cache'_j = \cache_j[x \mapsto \bot] 
            }{\cache,\mem \xrightarrow{\texttt{MStore}_i(x,v)} \cache',\mem'}
        $}
        \and
        \adjustbox{valign=t}{$
            \inferrule[\texttt{Crash}]{
            \cache'_i = \lambda x \ldotp \bot \\
            \forall j \neq i \ldotp \cache'_j = \cache_j \\\\
            \mem'_i = {\begin{cases}
            \lambda x \ldotp 0 & i \text{ has volatile memory} \\
            \mem_i & i \text{ has non-volatile memory}
            \end{cases}} \\\\
            \forall j \neq i \ldotp \mem'_j = \mem_j 
            }{\cache,\mem \xrightarrow{\mycrash{i}} \cache',\mem'}
        $}
    \end{mathpar}
    
    \vspace{-0.5em} 

    \begin{mathpar}
        \adjustbox{valign=t}{$
            \inferrule[\texttt{LOAD-from-C}]{
            \cache_j(x) = v \neq \bot \\\\
            \cache'_i = \cache_i[x \mapsto v] \\\\
            \forall j \neq i \ldotp \cache'_j = \cache_j 
            }{\cache,\mem \xrightarrow{\texttt{Load}_i(x,v)} \cache',\mem}
        $}
        \and
        \adjustbox{valign=t}{$
            \inferrule[\texttt{LOAD-from-M}]{
            \forall j \ldotp \cache_j(x) = \bot \\\\
            x \in \mathit{Loc}_k \\ \mem_k(x) = v
            }{\cache,\mem \xrightarrow{\texttt{Load}_i(x,v)} \cache,\mem}
        $}
        \and
        \adjustbox{valign=t}{$
            \inferrule[\texttt{Propagate-C-C}]{
            i \neq k \\
            x \in \mathit{Loc}_k \\
            \cache_i(x) = v \neq \bot  \\\\
            \cache'_i = \cache_i[x \mapsto \bot] \\\\
            \cache'_k = \cache_k[x \mapsto v] \\\\
            \forall j \not\in \{i,k\} \ldotp \cache'_j = \cache_j 
            }{\cache,\mem \xrightarrow{\tau} \cache',\mem}
        $}
        \and
        \adjustbox{valign=t}{$
            \inferrule[\texttt{Propagate-C-M}]{
            x \in \mathit{Loc}_k \\
            \cache_k(x) = v \neq \bot \\\\
            \forall j \ldotp \cache'_j = \cache_j[x \mapsto \bot] \\\\
            \mem'_k = \mem_k[x \mapsto v] \\\\
            \forall j \neq k \ldotp \mem'_j = \mem_j 
            }{\cache,\mem \xrightarrow{\tau} \cache',\mem'}
        $}
    \end{mathpar}

    \vspace{-0.5em}

    \begin{mathpar}
        \adjustbox{valign=t}{$
            \inferrule[\texttt{LFLUSH}]{
            \cache_i(x) = \bot
            }{\cache,\mem \xrightarrow{\texttt{LFlush}_i(x)} \cache,\mem}
        $}
        \and
        \adjustbox{valign=t}{$
            \inferrule[\texttt{RFLUSH}]{
            \forall j \ldotp \cache_j(x) = \bot
            }{\cache,\mem \xrightarrow{\texttt{RFlush}_i(x)} \cache,\mem}
        $}
        \and
        \adjustbox{valign=t}{$
            \inferrule[\texttt{Global-Persistent-Flush}]{ 
            \forall j,x \ldotp \cache_j(x) = \bot
            }{\cache,\mem \xrightarrow{\texttt{GPF}_i} \cache,\mem}
        $}
    \end{mathpar}
    \end{minipage}
}
\caption{\MODEL: Operational Semantics (Load, Store, Crash, and Flush Primitives)}
\label{fig:opsem}
\end{figure*}
}


\newcommand{\opsemRMW}{
\begin{figure*}[h]
\scalebox{0.9}{
\begin{mathpar}
\small
\inferrule[\texttt{L-Read-from-Cache-Modify-Write}]{
\cache_j(x) = v \neq \bot \\\\
\cache'_i = \cache_i[x \mapsto v'] \\\\
\forall j \neq i \ldotp \cache'_j = \cache_j[x \mapsto \bot] 
}{\cache,\mem \xrightarrow{\texttt{L-RMW}_i(x,v,v')} \cache',\mem}
\and
\inferrule[\texttt{L-Read-from-Mem-Modify-Write}]{
\forall j \ldotp \cache_j(x) = \bot \\\\
x \in \mathit{Loc}_k \\ \mem_k(x) = v \\\\
\cache'_i = \cache_i[x \mapsto v'] \\\\
\forall j \neq i \ldotp \cache'_j = \cache_j[x \mapsto \bot] 
}{\cache,\mem \xrightarrow{\texttt{L-RMW}_i(x,v,v')} \cache',\mem}
\and
\inferrule[\texttt{R-Read-from-Cache-Modify-Write}]{
x \in \mathit{Loc}_k \\\\
\cache_j(x) = v \neq \bot \\\\
\cache'_k = \cache_k[x \mapsto v'] \\\\
\forall j \neq k \ldotp \cache'_j = \cache_j[x \mapsto \bot] 
}{\cache,\mem \xrightarrow{\texttt{R-RMW}_i(x,v,v')} \cache',\mem}
\and
\inferrule[\texttt{R-Read-from-Mem-Modify-Write}]{
\forall j \ldotp \cache_j(x) = \bot \\\\
x \in \mathit{Loc}_k \\ \mem_k(x) = v \\\\
\cache'_k = \cache_k[x \mapsto v'] \\\\
\forall j \neq i \ldotp \cache'_j = \cache_j[x \mapsto \bot] 
}{\cache,\mem \xrightarrow{\texttt{R-RMW}_i(x,v,v')} \cache',\mem}
\and
\inferrule[\texttt{M-Read-from-Cache-Modify-Write}]{
x \in \mathit{Loc}_k \\\\
\cache_j(x) = v \neq \bot \\\\
\cache'_k = \cache_k[x \mapsto v'] \\\\
\forall j \neq k \ldotp \cache'_j = \cache_j[x \mapsto \bot] 
}{\cache,\mem \xrightarrow{\texttt{M-RMW}_i(x,v,v')} \cache',\mem}
\and
\inferrule[\texttt{M-Read-from-Mem-Modify-Write}]{
x \in \mathit{Loc}_k \\ \mem_k(x) = v \\\\
\forall j \ldotp \cache_j(x) = \bot \\\\
\mem'_k = \mem_k[x \mapsto v'] \\\\
\forall j \neq i \ldotp \cache'_j = \cache_j[x \mapsto \bot] 
}{\cache,\mem \xrightarrow{\texttt{M-RMW}_i(x,v,v')} \cache',\mem'} 
\end{mathpar}
}
\caption{\MODEL: Operational Semantics (Atomic Read-Modify-Write Primitives)}
\label{fig:opsem-rmw-atomics}
\end{figure*}
}

\section{A General Programming  Model for CXL}\label{sec:model}
In this section, we describe the most general system and failure model, and an abstract model for memory access on top of CXL. Based on these, we define operational semantics which provides us with a programming model for concurrent programs that access memory via the underlying CXL standard. To develop some intuition about valid behaviors of programs under our model, we provide a collection of litmus tests and prove various properties of the model. We present two variations of the programming model for specific hardware features.

\subsection{System model}

We consider an ideal model of a distributed system which consists of multiple Type 2 devices.
Each machine has computational capacity, memory capacity, or both. 
Memory accesses are cache-coherent.
Each memory address is mapped to a single physical volatile or persistent memory location on a specific machine. Memory accesses are performed at cache line granularity.
Machines are connected via a CXL switch and each machine manages the coherence of the memory attached to it. Machines send and receive CXL messages on dedicated ports. A system may incorporate several types of processing units (CPUs of different architectures, GPUs, FGPAs, etc.). Attached memory may also be of various types (e.g., SSD, DRAM, or NVMM). Some nodes may be only memory nodes, while others can host no shared memory at all. Deciding which parts of the memory are shared is up to the system designer. The CXL standard accommodates complex topologies and access control, so different memory segments located on a given machine may be shared with different subsystems simultaneously. Potentially, the entire address space may be shared in a cache-coherent manner. 

\remove{
\begin{figure}[t]
\centering
\includegraphics[scale=0.3, trim = 2cm 3cm 2cm 4cm, clip]{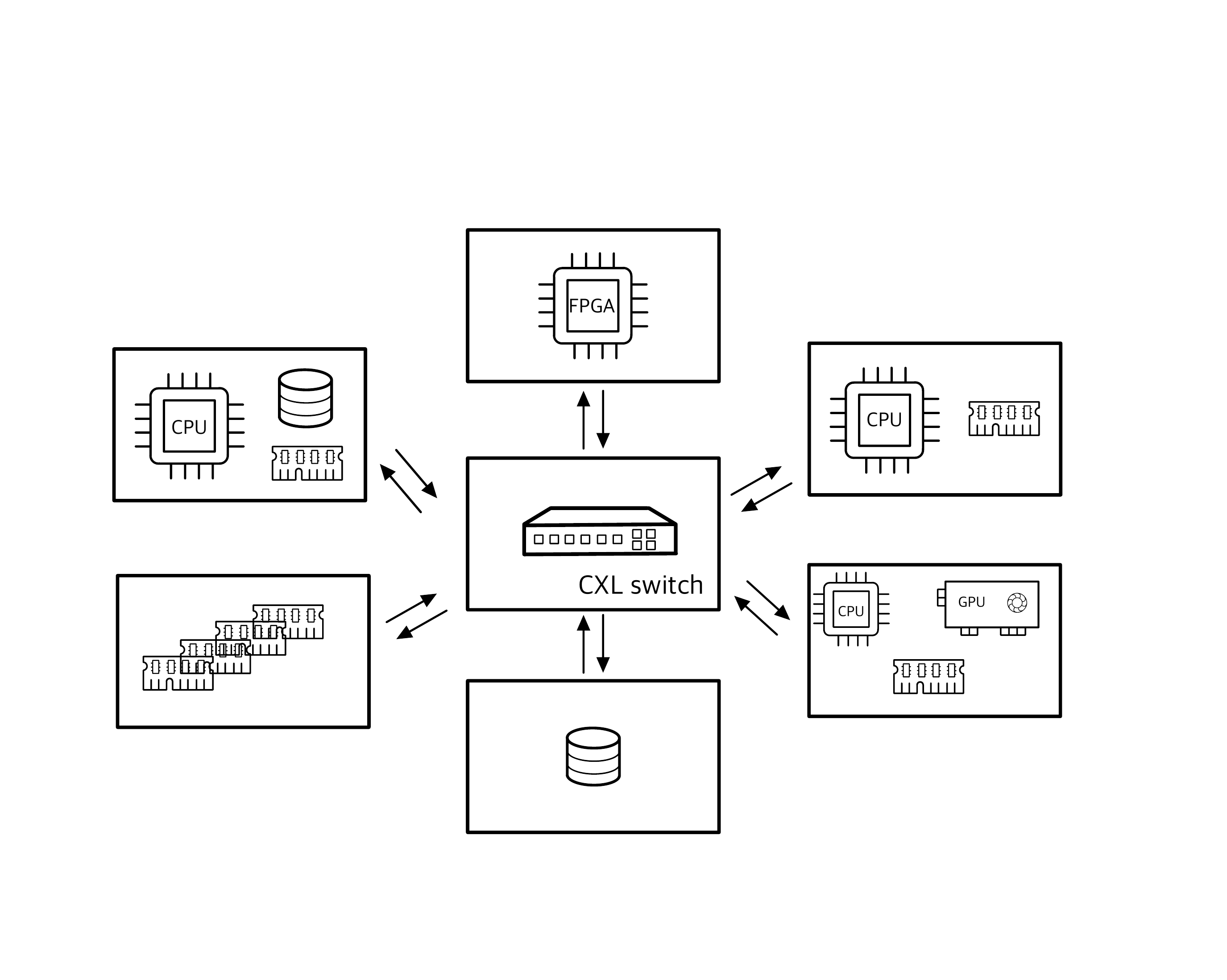} 
\caption{Multiple machines sharing memory connected through a CXL switch.}
\label{fig:cxl}
\end{figure}
}

We model each connected node as emitting CXL transactions (writes, reads, flushes) to the network in a specific order. Within each node, different consistency models may govern how instructions across cores propagate to the CXL network. The standard interconnects several machines, each potentially employing diverse underlying architectures with varying consistency guarantees. We do not consider the internal memory model of each machine, as the system is free to reorder messages within the CXL fabric (due to the properties of the underlying PCIe) \cite{sharma2023introduction}.
Hence, ordering has to be explicitly enforced when accessing remote memory locations, as the local memory ordering is unaware of the possibility of further reordering.



\paragraph{Failure model}
Our model permits spontaneous, independent crashes of any machine.
A machine crashes with its caches and the portion of shared memory attached to it.
We assume caches are volatile, so their contents vanish upon crash.
Data is considered lost upon crash unless it reached physical persistent media, or another failure domain before the crash.
When a machine crashes, the local state of any thread currently executing on it is lost as well (including program counter, registers, etc.).
Upon recovery, new threads are spawned to replace the crashed ones,
and all volatile memory is reset to some initial value.
The effects of store operations performed by the crashed machine that had not yet propagated into persistence or to another machine are lost.
Store operations to addresses belonging to the crashed machine may be lost if not persisted, or may also reside in another machine's cache and propagate at a later time. 

\subsection{Abstract CXL model}
The CXL \spec describes multiple types of read and write transactions as part of its sub-protocols. Such transactions refer to low-level transactions that are different from high-level, application ones. 
We provide an abstraction of these low-level transactions by mapping them to a single abstract read operation, three abstract store operations, and two abstract flush operations. These operations are referred to as \MODEL primitives in this work.
In \cref{sec:cxl0_mapping}, we provide a mapping from CXL transactions to \MODEL primitives.

\begin{figure}[t]
    \centering
    \includegraphics[width=1.1\columnwidth]{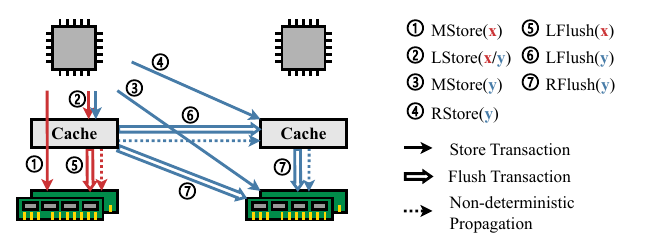}
    \caption{Semantics of \MODEL primitives as a layer of abstraction over CXL store and flush transactions (including non-deterministic propagation).}
    \label{fig:abstraction}
\end{figure}

\Cref{fig:abstraction} depicts the \MODEL primitives using a simple example. In this example, all primitives are performed by the left node and are labeled with numbers (\eg, \ding{192}). Continuous and double lines represent store and flush operations, respectively. Dotted lines represent eviction steps that are induced by the cache replacement policy. An arrowhead marks where the value is guaranteed to be propagated after the instruction or propagation step. Here, we assume that $x$ is a memory address allocated on the left node and $y$ on the right one.

Our model considers a single \texttt{Load} primitive.
The reason is that we assume cache coherence (i.e., reads-see-last-write), and this work aims at modeling high-level behaviors of concurrent programs rather than the coherence protocol itself. 

\noindent
The three abstract \MODEL store primitives are:
\begin{itemize}[leftmargin=*,itemsep=0pt]
    \item Local Store (henceforth \texttt{LStore}): A store to local cache. An \texttt{LStore} is considered complete when written to a local cache, and the CXL protocol does not provide any guarantees regarding when a local store is propagated to the physical memory where the stored address belongs. This store type does not generate any data transfer between machines and is likely to be preferable in terms of performance. As the \ding{193} label shows, an \texttt{LStore} has the same behavior when storing to each of the addresses ($x$ and $y$).
    \item Remote Store (henceforth \texttt{RStore}): A store to remote cache. An \texttt{RStore} is considered complete when it reaches either the address owner's cache or physical memory. CXL does not guarantee when a cache line is written back (namely, propagated) from the owner's cache to the physical memory. An \texttt{RStore} to $x$ executed by the left machine is equivalent to an \texttt{LStore}, whereas an \texttt{RStore} to $y$ (\ding{195}) results in writing the new value to the right machine's cache. 
    \item Memory Store (henceforth \texttt{MStore}): A store that completes only after reaching physical memory. The issuer may consider a write complete only after it has reached its final destination. If the destination memory is volatile and on the same failure domain, \texttt{RStore} and \texttt{MStore} are semantically equivalent under our failure model. In the example, an \texttt{MStore} to $x$ (\ding{192}) is considered complete only when the new value of $x$ is stored in main memory, and an \texttt{MStore} to $y$ (\ding{194}) results in writing to the physical memory of the right machine.
\end{itemize}

In addition to loads and stores, we consider two distinct \texttt{Flush} primitives, which are used to explicitly evict a cache line from a cache to the next level in the memory hierarchy.
Such write-back operations are common in commodity CPUs (\eg, \texttt{CLFLUSH} on x86~\cite{x86} and \texttt{DC-CVAC} on ARMv8~\cite{armv8}). CXL also introduces the notion of flushing a remote cache line (\texttt{CLFlush} in \cite{CXLSpec40}), namely, invalidating a cache line such that the cache controller writes back its contents to remote memory.
Our abstraction for flushes is as follows:
\begin{itemize}[leftmargin=*,itemsep=0pt]
    \item A Local Flush (henceforth \texttt{LFlush}) writes back a cache line that is stored locally (i.e., on the executing machine) to the next hierarchy, be it a remote cache or local main memory.  \texttt{LFlush} to $x$ results in updating the local memory (\ding{196}), and the effect 
 of \texttt{LFlush} to $y$ is an update to the right machine's cache (\ding{197}).
    \item A Remote Flush (henceforth \texttt{RFlush}) writes back a cache line to the corresponding main memory. \texttt{RFlush} to $y$ causes a write back from either cache to the right machine's physical memory $y$ (\ding{198}), and \texttt{RFlush} to $x$ is equivalent to \texttt{LFlush} to $x$ in this example.
\end{itemize}


\remove{
\texttt{LFlush}-like instructions, which flush a cache line from cache to memory, exist in many CPU architectures (\eg, x86's \texttt{CLFlush}). However, to the best of our knowledge, commodity CPU architectures do not include \texttt{RFlush}-equivalent instructions, as invalidating a remote cache line had not been a relevant operation before CXL.
\ori{last para seems redundant to me. Quite repetitive to the text before the items.}}

The \texttt{Global Persistent Flush (GPF)} instruction is also supported by the \spec (\S9.8 in \cite{CXLSpec40}).
It initiates a 2-phase write-back protocol to drain \emph{all} caches of all CXL agents in the cache coherence domain.
We regard it as rather complex and highly susceptible to additional machine failures.
Moreover, the specification treats GPF as optional and explicitly omits a list of triggering events.
Therefore, we capture the behavior of \texttt{GPF} in our model for completeness, but we argue that it does not reliably provide the same persistence guarantees as, for example, Intel's eADR technology~\cite{intel2021eadr} does in a single machine setting.

Using a \texttt{GPF} in a weakly ordered memory model exposes the system to potential inconsistencies since the order by which writes reach persistent media might still differ from the program order. In this case, a value of a write operation could already reside in a cache while an earlier write by the same thread still resides in the processor's internal buffers. The persistent state is guaranteed to be consistent only if the earlier write's effects are applied to the persistent state before the later write's. This cannot be guaranteed solely by using persistent caches or \texttt{GPF}, in presence of volatile buffers in processors that do not provide total store order (TSO).
\texttt{GPF} is intended to be used for a planned shutdown or reboot of the system. Therefore, we do not expect it to be used in application level algorithms. However, a carefully designed algorithm may still employ \texttt{GPF} for snapshots, thanks to its global and blocking properties.

\remove{
The \texttt{LFlush} operation resembles instructions like x86's \texttt{CLFlush} (which flushes a cache line from cache to memory), whereas to our knowledge \texttt{RFlush} does not have an equivalent instruction in the context of a single machine, due to the fact that invalidating a remote cache line had not been a relevant operation before CXL. 
}


Standard cache replacement policies introduce nondeterministic (from the perspective of the program) evictions of cache lines from a cache to the next memory hierarchy. This is not driven by any instruction and is considered normal behavior by the CXL \spec. These nondeterministic steps are captured by the dotted lines: propagation is possible from a local cache to a remote cache or local memory. Propagation from local cache to remote memory is achieved by these two steps taking place one after the other.

\paragraph{Limitations of CXL}
Under the current specification, we identify three key limitations related to concurrent programming.
First, the CXL \spec lacks ordering primitives.
It defines neither global ordering primitives (like the \lstinline{__threadfence_system} call in CUDA~\cite{CUDA}, which is a GPU-wide fence), nor local ones (like x86's \texttt{MFENCE}~\cite{x86}).
Consequently, our model does not have memory fences.
Second, the CXL \spec only specifies synchronous flushes, which take effect immediately, and does not provide asynchronous flush instructions, whose effect is delayed until a corresponding barrier.
Such instructions exist on multicore architectures:
\texttt{CLFLUSHOPT} and \texttt{CLWB}  followed by an \texttt{SFENCE} barrier on x86, and 
\texttt{DC.CVAP} followed by a \texttt{DSB.SY} barrier on ARM.
These allow fine-tuned efficient implementations, allowing programmers to mark certain stores that are needed to persist, and later make sure they all persist by one fence. Adding such features to \MODEL can be done along the lines of \cite{RaadWNV20, KhyzhaL21} using an additional layer of partially ordered persistency buffers, or following \cite{ChoLRK21} with view-based semantics.

\remove{
\paragraph{CXL on existing architectures}
Dealing with a technology that is not yet fully supported, we make a general assumption that a mapping from CXL transactions to higher-level languages will be available, based on the notion that the programmer must be able to determine the type of access to different memory types over CXL. 
}

\subsection{Formal model}
We now introduce \MODEL, the first programming model for CXL.
We formally model the system as a labeled transition system (LTS).
We assume $N$ machines communicating through CXL. 
Each machine emits actions in an order that is determined by the program order and the memory model of its local compute unit.
The set of memory locations on machine $i$ is denoted by $\mathit{Loc}_i$, and we assume that $\mathit{Loc}_1, \ldots, \mathit{Loc}_N$ are pairwise disjoint.
We let $\mathit{Loc} = \cup_{i}\mathit{Loc}_i$ denote the set of all shared locations in the system.
Values are taken from a set $\mathit{Val}$, which we assume to include a distinguished value, denoted by $0$, used for the initial value of every location.

\noindent
The transition labels in \MODEL consist of: 

\begin{itemize}[leftmargin=*,itemsep=2pt]
    \item Labels for actions emitted by the networked machines:
        $\texttt{LStore}_i(x,v)$,
        $\texttt{RStore}_i(x,v)$,
        $\texttt{MStore}_i(x,v)$,
        $\texttt{Load}_i(x,v)$,\\
        $\texttt{LFlush}_i(x)$, 
        $\texttt{RFlush}_i(x)$,
        $\texttt{GPF}_i$, and
        $\texttt{RMW}_i(x, old, new)$\\
        where $x\in \mathit{Loc}$ and $v,old,new\in \mathit{Val}$.
    \item Silent nondeterministic internal propagation $\tau$.
    \item Nondeterministic local crash $\mycrash{i}$
        where $i\in \{1, \ldots, N\}$.
\end{itemize}

\noindent
In turn, \MODEL system states are pairs $\conf=(\cache,\mem)$, where:

\begin{itemize}[leftmargin=*,itemsep=2pt]
    \item $\cache$ maps each machine $i$ to its \emph{local cache}\\
        $\cache_i: \mathit{Loc}\to \mathit{Val}\uplus\{\bot\}$.
    \item $\mem$ maps each machine $i$ to its \emph{local memory}\\
        $\mem_i: \mathit{Loc}_i \to \mathit{Val}$.
\end{itemize}

For every machine $i$, $\cache_i$ and $\mem_i$ are functions from the address space to possible values.
$\cache_i$ may additionally receive the invalid value, denoted by $\bot$.
$\mem_i$ maps addresses belonging to machine $i$ to valid values, and represents the values as they are stored in its physical memory.
For brevity, we assume that for each $i$, $\mem_i$ is either \emph{volatile} or \emph{non-volatile}, which determines whether or not it loses its contents upon crash.
Expanding the model to support machines with both volatile and non-volatile memory is simple using sub-indices.

The abstract components ``cache'' and ``memory'' mentioned above capture how far the latest value of an address propagated towards physical memory in the memory hierarchy.
Although we refer to them as ``cache'' and ``memory'', they do \emph{not} directly correspond to the hardware primitives with the same names.
Our model merely captures how the effects of write operations on shared memory propagate and become persistent.
While the CXL standard presents both cacheable and non-cacheable write transactions, \MODEL is agnostic about the local cacheability of reads and writes.
It assumes the underlying cache coherence and does not provide any means to reason about it. 

\medskip
The following invariant holds throughout executions:
\begin{gather*}
\footnotesize
\forall i, j \in \{1, \ldots, N\} \ldotp \forall x \in \mathit{Loc} \ldotp \\
\cache_i(x)\neq\bot \land \cache_j(x)\neq\bot \implies \cache_i(x)=\cache_j(x)
\end{gather*}

This invariant asserts that only one unique valid value may be stored for a given logical address in all caches across the system.
Roughly speaking, this is similar to a cache line being in a \textsc{shared} state for some coherence protocol. 
However, as noted above, our model does not aim to provide a precise description of CXL's coherence state machine.
Instead, we abstract it away, only describing the behavior of concurrent program executions.
It is possible to have a configuration in which the value of a location $x$ in caches differs from the value in the memory of the machine that owns $x$.
This is necessary to model the fact that writes become globally visible before they are persistent.
In such cases, the cache value is always more recent.

\adjopsem


Initially, all machines start with empty caches ($\cache_i = \lambda x \ldotp \bot$, meaning every value of $x$ is mapped to $\bot$) and zero-initialized memories ($\mem_i = \lambda x \ldotp 0$).
\Cref{fig:opsem} presents the steps of the operational semantics of the system. 

\paragraph{Store steps}
For a location $x$ that belongs to machine $k$, 
$\texttt{LStore}_i(x,v)$ sets $x$ to $v$ in $\cache_i$;
$\texttt{RStore}_i(x,v)$ sets $x$ to $v$ in $\cache_k$;
and $\texttt{MStore}_i(x,v)$
sets $x$ to $v$ in $\mem_k$.
All these steps also invalidate $x$ in all other caches to make sure the previous value is not available anywhere.

\paragraph{Load steps}
If a value exists in some cache, then $\texttt{Load}_i(x,v)$ takes its value $v$ from there.
The global invariant ensures that valid values in all caches are the same.
In addition, a load from cache copies the value to the cache of machine $i$.
The latter ensures the correct behavior of a future $\texttt{LFlush}$ operation.
If no cache has a value for $x$, the value is loaded from the memory that contains address $x$.

\paragraph{Propagation steps}
The system propagates values nondeterministically between caches (``horizontal propagation'') and from cache to memory (``vertical propagation'').
Cache-to-cache propagation of a value for address $x$ from machine $i$ is always toward the machine that owns $x$ and removes the value from the machine $i$'s cache.
Cache-to-memory propagation is always from the owner's cache to the owner's memory and removes the value from all caches.

\paragraph{Flush steps}
Intuitively, $\texttt{LFlush}_i(x)$ forces the horizontal propagation of $x$'s value from $\cache_i$ to the cache of its owner. 
When performed by the owner of $x$, $\texttt{LFlush}_i(x)$ forces vertical propagation to memory.
In the presence of nondeterministic propagation steps, we achieve this effect by ensuring that such propagation has already occurred through a precondition ($\cache_i(x) = \bot$) of $\texttt{LFlush}_i(x)$.
(This is similar to the modeling of $\texttt{MFENCE}$ in TSO, where, instead of actively draining the store buffer, the fence simply blocks until the buffer is drained by the silent nondeterministic propagation steps \cite{RaadWNV20}.)
Similarly, $\texttt{RFlush}_i(x)$ forces the full propagation to the owner's memory by ``blocking'' until the value does not exist in any of the caches.
For \texttt{GPF}, all data that resides in caches must be propagated to the memory on which it is allocated.
This step forces both horizontal and vertical propagation, and behaves like an \texttt{RFlush} over the entire address space.
We model this step in the same blocking manner.

\paragraph{Crash step}
At any point during execution, a machine may crash non-deterministically.
This results in the machine losing the contents of its cache, and, if its memory is volatile, re-initializing the memory to zero.

\paragraph{Read-modify-write Operations}
The CXL specification states that: ``\textit{Each host can perform arbitrary atomic operations supported by its Instruction-Set Architecture (ISA) by gaining \textsc{Exclusive} access to a cache line, \emph{[and]} performing the atomic operation on it within its cache}'' (\S 2.4.4 in \cite{CXLSpec40}).
Therefore, we model read-modify-write (RMW) operations, such as compare-and-swap (CAS) and fetch-and-add (FAA), as an atomic combination of a load followed by a store without allowing interference between the two.
This is similar to how they are specified in standard operational presentation of a strongly consistent shared-memory model.
Given that we have three types of store instructions, \MODEL essentially includes \texttt{L-RMW}, \texttt{R-RMW}, and \texttt{M-RMW} primitives.

The operational semantics of the RMW operations are a straightforward extension of the existing semantics, and for brevity, we only depict \texttt{L-RMW}:
\begin{mathpar}
\small
\inferrule*{
\cache_j(x) = v \neq \bot \\\\
\cache'_i = \cache_i[x \mapsto v'] \\\\
\forall j \neq i \ldotp \cache'_j = \cache_j[x \mapsto \bot] 
}{\cache,\mem \xrightarrow{\texttt{L-RMW}_i(x,v,v')} \cache',\mem}
\and 
\inferrule*{
\forall j \ldotp \cache_j(x) = \bot \\\\
x \in \mathit{Loc}_k \\ \mem_k(x) = v \\\\
\cache'_i = \cache_i[x \mapsto v'] \\\\
\forall j \neq i \ldotp \cache'_j = \cache_j[x \mapsto \bot] 
}{\cache,\mem \xrightarrow{\texttt{L-RMW}_i(x,v,v')} \cache',\mem}
\end{mathpar}



We remind that these steps are combinations of the load and store steps, and hence it is required to distinguish between loads from $\cache$ and $\mem$, as well as the different types of store operations in \MODEL.
This results in a total of 6 different RMW operations.
Note also that a failed RMW, for example, a failed CAS, is equivalent to a plain read.
Therefore, the citation from the CXL specification describes only our \texttt{L-RMW}, but implementing the other RMW operations is possible if the cache line is kept in \textsc{Exclusive} or \textsc{Modified} state until the post-conditions for the remote store are met.


We note that the intricacies of the model arise from the need to handle node crashes.
Without crashes, \MODEL has simple, sequentially consistent semantics.
Every load reads the last written value to the same location, flush instructions are benign, and all store instructions are equivalent and executed in program order.


\subsection{Intuition}

Before demonstrating the behaviors of \MODEL in a few examples, we highlight several useful (and expected) implications.
We write
{\footnotesize $\conf \xrightarrow{\alpha_1 \ldots \alpha_n} \conf'$}
if there is a sequence of transitions labeled
$\alpha_1, \ldots, \alpha_n$,
possibly interleaved with additional silent $\tau$-steps, starting in state $\conf$ and ending in state $\conf'$.


\setlength{\jot}{0pt}
\begin{proposition}
\label{prop:one}
The following hold assuming $x \in \mathit{Loc}_k$; $j\neq k$:
\begin{enumerate}[leftmargin=*,itemsep=4pt]
\item \texttt{RStore} is stronger than \texttt{LStore}:\\
If $\conf \xrightarrow{\texttt{RStore}_i(x,v)} \conf'$,
then $\conf \xrightarrow{\texttt{LStore}_i(x,v)} \conf'$.
\item \texttt{RStore} and \texttt{LStore} by the owner are equivalent:\\
If $\conf \xrightarrow{\texttt{LStore}_k(x,v)} \conf'$,
then $\conf \xrightarrow{\texttt{RStore}_k(x,v)} \conf'$.
\item \texttt{MStore} is stronger than \texttt{RStore}:\\
If $\conf \xrightarrow{\texttt{MStore}_i(x,v)} \conf'$,
then $\conf \xrightarrow{\texttt{RStore}_i(x,v)} \conf'$.
\item \texttt{RFlush} is stronger than \texttt{LFlush}:\\
If $\conf \xrightarrow{\texttt{RFlush}_i(x)} \conf'$,
then $\conf \xrightarrow{\texttt{LFlush}_i(x)} \conf'$.
\item \texttt{LFlush} after \texttt{RStore} by non-owner is redundant:\\
If $\conf \xrightarrow{\texttt{RStore}_j(x,v)} \conf'$,
then $\conf \xrightarrow{\texttt{RStore}_j(x,v) \cdot \texttt{LFlush}_j(x)} \conf'$.
\item \texttt{RFlush} after \texttt{MStore} is redundant:\\
If $\conf \xrightarrow{\texttt{MStore}_i(x,v)} \conf'$,
then $\conf \xrightarrow{\texttt{MStore}_i(x,v) \cdot \texttt{RFlush}_i(x)} \conf'$.
\item \texttt{RStore} by non-owner is simulated by \texttt{LStore} + \texttt{LFlush}:\\
If $\conf \xrightarrow{\texttt{LStore}_j(x,v) \cdot \texttt{LFlush}_j(x)} \conf'$,
then $\conf \xrightarrow{\texttt{RStore}_j(x,v)} \conf'$.
\item \texttt{MStore} is simulated by \texttt{LStore} + \texttt{RFlush}:\\
If $\conf \xrightarrow{\texttt{LStore}_i(x,v) \cdot \texttt{RFlush}_i(x)} \conf'$,
then $\conf \xrightarrow{\texttt{MStore}_i(x,v)} \conf'$. \label{inprop:MstoreLstore}
\end{enumerate}
\end{proposition}

\medskip
To establish confidence, we have formulated and proved \cref{prop:one} in Rocq.
The Rocq files are available in the supplementary material.
Most of the arguments are straightforward.
\cref{sec:prop_proof} of the supplementary material contains proofs for $(7)$ and $(8)$, which are more difficult to prove. 

\mytext{
For $(7)$, we first observe that it suffices to show that $\conf_1 \xrightarrow{\texttt{LStore}_j(x,v)} \conf_2$, 
$\conf_2 \xRightarrow{} \conf_3$ (with a trace that consists solely of internal propagation steps),
and $\conf_3 \xrightarrow{\texttt{LFlush}_j(x)} \conf_4$ together imply that
$\conf_1 \xRightarrow{\texttt{RStore}_j(x,v)} \conf_4$.
The proof of this claim proceeds by induction on the length of the trace from $\conf_2$ to $\conf_3$.
If this trace starts with a propagation (either \texttt{Cache-Cache} or \texttt{Cache-Mem}) of a location $y$ different from $x$, we can commute this propagation before the \texttt{LStore}, and derive $\conf_1 \xRightarrow{\texttt{RStore}_j(x,v)} \conf_4$ from the induction hypothesis. If this trace starts with a propagation of $x$, it must be a \texttt{Cache-Cache} propagation in node $i$. Then, the effect of $\texttt{LStore}$ followed by this propagation is similar to the effect of $\texttt{RStore}$. Finally, we observe that $\conf_3 \xrightarrow{\texttt{LFlush}_j(x)} \conf_4$ ensures that the analysed trace must contain a propagation step of $x$.

Item $(8)$ is proved similarly, but requires another nested induction. 
Again we observe that it suffices to show that $\conf_1 \xrightarrow{\texttt{LStore}_j(x,v)} \conf_2$, 
$\conf_2 \xRightarrow{} \conf_3$,
and $\conf_3 \xrightarrow{\texttt{RFlush}_j(x)} \conf_4$ together imply that
$\conf_1 \xRightarrow{\texttt{MStore}_j(x,v)} \conf_4$.
We prove this claim by an induction similar to the one above.
For the case that the middle trace of propagations starts with a \texttt{Cache-Cache} propagation of $x$, we observe that the effect of this propagation after the $\texttt{LStore}$ is identical to an $\texttt{RStore}$, and use another (similar) induction argument to show that
$\conf_1 \xrightarrow{\texttt{RStore}_j(x,v)} \conf_2$, 
$\conf_2 \xRightarrow{} \conf_3$,
and $\conf_3 \xrightarrow{\texttt{RFlush}_j(x)} \conf_4$ together imply that
$\conf_1 \xRightarrow{\texttt{MStore}_j(x,v)} \conf_4$.
}

Next, we discuss possible and impossible behaviors under the \MODEL model through a set of litmus tests.
We find that some of the possible behaviors are undesirable and raise consistency issues in the face of a crash. 
Figure \ref{fig:litmus-abs} depicts the litmus tests as sequences of events and operations as they are performed by the local machine in \emph{execution order}. 
We assume that the execution order may differ from the program order, as is common in commodity CPUs.
We denote memory allocated on machine $i$ as $x^i$, and we assume that all memory in the following tests is non-volatile.
Note that $\texttt{RStore}_2(y^1, x^2)$ is a shorthand for reading $x^2$ into a local register and then performing an \texttt{RStore}
to $y^1$.

\setlength{\jot}{0pt}
\begin{figure}[htb]
\begin{minipage}{\columnwidth}
\footnotesize
\begin{align}
&\texttt{RStore}_1(x^1,1);\allowbreak\ \text{\Lightning}_1;\allowbreak\ \texttt{Load}_1(x^1,0) &\text{\mycheck} \label{eq:lit-1}\\
&\texttt{MStore}_1(x^1,1);\allowbreak\ \text{\Lightning}_1;\allowbreak\ \texttt{Load}_1(x^1,0) &\text{\myxmark} \label{eq:lit-2}\\
&\texttt{LStore}_1(x^1,1);\allowbreak\ \texttt{LFlush}_1(x^1);\allowbreak\ \text{\Lightning}_1;\allowbreak\ \texttt{Load}_1(x^1,0) &\text{\myxmark} \label{eq:lit-3}\\
&\texttt{LStore}_1(x^2,1);\allowbreak\ \texttt{LFlush}_1(x^2);\allowbreak\ \mycrash{2};\allowbreak\ \texttt{Load}_1(x^2,0) &\text{\mycheck} \label{eq:lit-4}\\
&\texttt{LStore}_1(x^2,1);\allowbreak\ \texttt{RFlush}_1(x^2);\allowbreak\ \mycrash{2};\allowbreak\ \texttt{Load}_1(x^2,0) &\text{\myxmark} \label{eq:lit-5}\\
&\texttt{LStore}_1(x^3,1);\allowbreak\ \texttt{Load}_2(x^3,1);\allowbreak\ \mycrash{1};\allowbreak\ \texttt{Load}_2(x^3,0) &\text{\myxmark} \label{eq:lit-6}\\
&\texttt{LStore}_1(x^3,1);\allowbreak\ \texttt{Load}_2(x^3,1);\allowbreak\ \texttt{LFlush}_2(x^3);\allowbreak\ \mycrash{1};\allowbreak\ \mycrash{2};\allowbreak\ \texttt{Load}_2(x^3,0) &\text{\myxmark} \label{eq:lit-7}\\
&\texttt{RStore}_1(x^2,1);\allowbreak\ \texttt{RStore}_2(y^1,x^2);\allowbreak\ \mycrash{2};\allowbreak\ \texttt{Load}_1(y^1,1);\allowbreak\ \texttt{Load}_1(x^2,0) &\text{\mycheck} \label{eq:lit-8}\\
&\texttt{MStore}_1(x^2,1);\allowbreak\ \texttt{RStore}_2(y^1,x^2);\allowbreak\ \mycrash{2};\allowbreak\ \texttt{Load}_1(y^1,1);\allowbreak\ \texttt{Load}_1(x^2,0) &\text{\myxmark} \label{eq:lit-9}
\end{align}
\end{minipage}
\caption{Litmus tests for \MODEL}
\label{fig:litmus-abs}
\end{figure}

We present the litmus tests as traces of \MODEL primitives. 
Although the CXL specification allows transactions operating on different addresses to execute concurrently, we serialize them in some possible order for simplicity.
We use a sequential presentation to determine the behavior of retired instructions, after they have been internally ordered by the processors executing them.
This also enables us to separate the discussion from any processor-specific memory model. 
Allowed behaviors are marked with \mycheck, and illegal behaviors are marked with \myxmark. We present scenarios that highlight the similarities and key differences between \MODEL and previous models that consider persistent memory.

Tests \ref{eq:lit-1}-\ref{eq:lit-3} span a single machine.
In the first test, the model permits the loss of a stored value upon crashing when using \texttt{RStore} because it does not guarantee that the value has propagated to persistence before the crash.
However, such behavior is impossible in test \ref{eq:lit-2} thanks to \texttt{MStore}, which guarantees the persistence of the update before it returns.
Test \ref{eq:lit-3} shows that a value cannot be lost if the store has been flushed (i.e., propagated) to the local persistent memory using \texttt{LFlush} before the crash. 

Tests \ref{eq:lit-4}-\ref{eq:lit-7} involve multiple machines.
In Test \ref{eq:lit-4}, a stored value may be lost due to a crash if it has not reached remote persistent memory.
Test \ref{eq:lit-5} uses the stronger \texttt{RFlush} which requires the value to propagate further (by requiring $\forall j \ldotp \cache_j = \bot$), thus preventing the loss of the stored value.
Tests \ref{eq:lit-6} and \ref{eq:lit-7} demonstrate that copying a value to the local cache upon loading helps prevent the loss of stored values when a machine performing an \texttt{LStore} to remote memory crashes.
A load operation reads the value in Test \ref{eq:lit-6} from $\cache_2$ after the crash, and in Test \ref{eq:lit-7} from $\cache_3$ after the crash, thanks to the flush operation by machine 2.

The last two tests concern writes to multiple variables.
Test \ref{eq:lit-8} shows that it is possible to lose a stored value that another operation had already observed.
Specifically, a possible recovered state would include the effects of a later operation without the first.
In that case, using \texttt{MStore} for the first write would make it impossible to obtain an inconsistent state upon recovery of the second machine, as seen in Test \ref{eq:lit-9}.


\subsection{Model Variants}
\label{sec:variants}

We describe two possible variants of our model to capture potential hardware implementation alternatives.

\paragraph{Crash with cache line poisoning}
Cache line poisoning marks a line as unusable after corruption (e.g., uncorrectable ECC errors) to prevent access to it.
CXL supports cache line poisoning upon crash of a machine.
In the event of a surprise hot unplug, link down, or transaction timeout where the device is unresponsive, the system uses CXL Isolation (\S9.9, \S12.3 in \cite{CXLSpec40}).
A Non-Existent Memory (MemData-NXM) response is generated with the poison bit set for read requests targeting the unreachable memory addresses, if the Poison on Decode Error capability is enabled (\S8.2.4, Table 8-117 in \cite{CXLSpec40}).
To capture this behavior, we provide the following alternative \texttt{Crash} step:
$$
\small
\inferrule[\texttt{Crash} (PSN)]{
\cache'_i = \lambda x \ldotp \bot \\
\forall j \neq i, x \not\in \mathit{Loc}_i \ldotp \cache'_j(x) = \cache_j(x) \\
\forall j \neq i, x \in \mathit{Loc}_i \ldotp \cache'_j(x) = \bot \\\\
\mem'_i = {\begin{cases}
\lambda x \ldotp 0 & i \text{ has volatile memory} \\
\mem_i & i \text{ has non-volatile memory}
\end{cases}} \\
\forall j \neq i \ldotp \mem'_j = \mem_j 
}{\cache,\mem \xrightarrow{\mycrash{i}} \cache',\mem'}
$$
After a crash, the crashed machine's addresses are mapped to $\bot$ in all caches.
We call this version of the model $\MODEL^{\text{PSN}}$. 




\remove{Viral Signaling: If a device detects a fatal uncorrectable error but remains operational, it conveys Viral status to the remote entity to contain the error. This is implemented by forcing a CRC error on outgoing flits and embedding viral status information in the retry acknowledgment message (\S4.2.9, \S12.4). \ga{we might want to ignore this one}}

\paragraph{Remote loads with implicit write-back}

Load steps in \MODEL are deterministic.
We do not model details like hardware write-back responsibility because the CXL \spec does not explicitly require remote loads to update memory.
However, if hardware turns out to implement such a policy, this would affect observable persistence and we can update our semantics to reflect this behavior.
To capture implicit write-backs for remote loads, i.e. every change is written to main memory before being propagated to another cache, we may replace our \texttt{LOAD-from-C} step by the following:
$$\small
    \inferrule*[left=\texttt{LOAD-from-C} \text{(LWB)}]{
        \cache_i(x) = v \neq \bot 
    }{\cache,\mem \xrightarrow{\texttt{Load}_i(x,v)} \cache,\mem}
$$
Intuitively, this forces that all loads for non-local addresses are served from memory.
We use the same trick from before to force the full propagation to the owner’s memory by “blocking” until the value does not exist in any of the caches.
We call this version of the model $\MODEL^{\text{LWB}}$.

\paragraph{Comparison of the different variants}

Every trace allowed by the above variants is also allowed by \MODEL.
We encode all model variants as communicating sequential processes, and use the FDR4 refinement checker~\cite{fdr3} to obtain examples showing traces of \MODEL that are not allowed by the variants, as well as examples showing that the two 
variants are incomparable.

For that matter, consider two machines: machine $1$ with NVMM and machine $2$ with volatile memory.
We denote the locations on machine $i$ as $x^i$ and report the validity (\mycheck/\myxmark) as a triple (\MODEL, $\MODEL^{\text{LWB}}$, $\MODEL^{\text{PSN}}$):

\setlength{\jot}{0pt}
\begin{minipage}{\columnwidth}
\footnotesize
\begin{align}
&\texttt{RStore}_2(x^1,1);\allowbreak\ \texttt{Load}_2(x^1,1);\allowbreak\ \text{\Lightning}_1;\allowbreak\ \texttt{Load}_2(x^1,0) &\text{\mycheck}, \text{\myxmark}, \text{\mycheck} \label{eq:lit-10}\\
&\texttt{LStore}_1(x^1,1);\allowbreak\ \texttt{Load}_2(x^1,1);\allowbreak\ \text{\Lightning}_1;\allowbreak\ \texttt{Load}_1(x^1,0) &\text{\mycheck}, \text{\myxmark}, \text{\mycheck} \label{eq:lit-11}\\
&\texttt{LStore}_2(x^1,1);\allowbreak\ \text{\Lightning}_1;\allowbreak\ \texttt{Load}_1(x^1,1);\allowbreak\ \text{\Lightning}_1;\allowbreak\ \texttt{Load}_2(x^1,0) &\text{\mycheck}, \text{\mycheck}, \text{\myxmark} \label{eq:lit-12}
\end{align}
\end{minipage}
\vspace{2pt}

In Test~\ref{eq:lit-10}, machine $2$ updates $x$ in the cache of machine $1$ and then loads $x$ into its own cache. After a crash of machine $1$, machine $2$ reads the old value of $x$ under both \MODEL and $\MODEL^{\text{PSN}}$. The same happens when the initial \texttt{RStore} of machine $2$ is replaced by an \texttt{LStore} of machine $1$ in Test~\ref{eq:lit-11}. This happens because both \MODEL and $\MODEL^{\text{PSN}}$ allow updates issued by \texttt{LStore} or \texttt{RStore} to be lost, as in both models loads may be served from the cache before the update becomes persistent. In particular, the value may still be propagated to the cache of machine $1$ after the load. Tests \ref{eq:lit-10} and \ref{eq:lit-11} illustrate that $\MODEL^{\text{LWB}}$ is the most resilient model with respect to losing updates.

In Test~\ref{eq:lit-12}, machine $2$ updates $x$ in its own cache, after which machine $1$ crashes. Subsequently, machine $1$ loads the latest value of $x$ and then crashes again. Finally, machine $2$ loads the old value of $x$. This test shows that $\MODEL^{\text{PSN}}$ prevents inconsistencies across consecutive crashes.
In contrast, both \MODEL and $\MODEL^{\text{LWB}}$ allow such behavior, as remotely cached updates must propagate back to memory through a set of volatile caches which are susceptible to crashes.
Poison semantics undercut this chain by poisoning of cache entries immediately upon the first crash.
Note that $\MODEL^{\text{LWB}}$ only writes back data to memory in the event of a remote load. Therefore, ${Load}_1(x^1,1)$ does not guarantee the data reaches main memory before the load finishes.


\section{System Model Variations}
\label{sec:newmodel}


As with any new standard, the adoption of CXL will be gradual as the hardware, software, and standard evolve.
In this section, we present a roadmap of viable system configurations based on discussions with experts from CXL vendors and on existing setups.
We characterize each configuration and demonstrate \MODEL's applicability to each.
\begin{figure}[t]
    \centering
    \begin{subfigure}[b]{0.5\columnwidth}
        \centering
        \includegraphics[keepaspectratio, width=\textwidth]{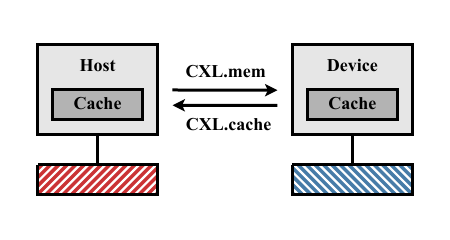}
        \caption{Host-Device Pairing}
        \label{fig:host-device}
    \end{subfigure}%
    \hfill
    \begin{subfigure}[b]{0.5\columnwidth}
        \centering
        \includegraphics[keepaspectratio, width=\textwidth]{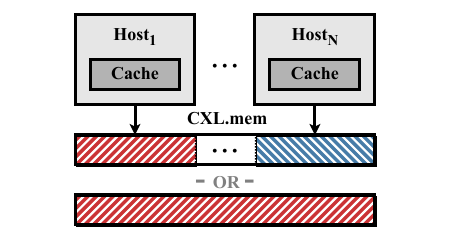}
        \caption{Memory Pool}
        \label{fig:mem-pool}
    \end{subfigure}%
    \hfill
    \caption{Current System Models with CXL}
    \label{fig:main}
\end{figure}



\paragraph{Host-device pair}
We begin with a simple setting consisting of a host and an accelerator that share memory in a cache-coherent manner using CXL.cache and CXL.mem.
It is depicted in \cref{fig:host-device}.
The host manages its own memory's coherence, while device memory coherence is either host-managed or bypassed, depending on the bias mode.
If memory is volatile, failure of the host or device may cause the loss of the attached memory's content.
This is the first step in the evolution of CXL deployments, as this single-machine setup has already been deployed \cite{demystify-cxl-type2-2024, accelerating-rag-2025}.

\MODEL applies to this setting with a few restrictions.
Both the host and the device are modeled as machines with their own cache and memory.
According to the CXL \spec, the host can issue all available \MODEL primitives apart from \texttt{RStore}, \texttt{LFlush} and remote RMWs (\texttt{R-RMW} and \texttt{M-RMW}).
The device can issue all stores, including \texttt{RStore}, but cannot issue \texttt{LFlush} and remote RMWs.
Note that \MODEL applies to device memory in host-bias mode.

\paragraph{Partitioned disaggregated memory pool}
Next, we discuss a setting in which multiple hosts access disjoint parts of a memory pool for memory extension.
It is depicted in \cref{fig:mem-pool}.
Memory is only shared between threads per host, therefore we only have intra-host memory coherence.
CXL.mem is the key protocol in use.
This realization of a disaggregated memory pool via CXL already exists \cite{pond, intel-flat-memory-mode-2024}.
It is conceptually similar to having a set of isolated machines with NVMM.
The memory pool acts as an external failure domain to the hosts.
Machine crashes do not affect the contents of the pool.

\MODEL applies to this setting with some restrictions.
The disjoint memory partitions are modeled as N additional memory nodes.
Further, there is a bijection from compute nodes to memory nodes, and no other memory is shared in the system.
Per the CXL \spec, we apply the following restrictions to the available \MODEL primitives.
We exclude \texttt{RStore}, \texttt{LOAD-from-C}, \texttt{Propagate-C-C}, and remote \texttt{RMW}s, as there is no interaction between hosts.
Additionally, we point out that \texttt{LFlush} and \texttt{RFlush} are semantically equivalent in this setting.
The remaining \MODEL primitives suffice to reason about any concurrent program execution interleaved with a crash.

\paragraph{Shared disaggregated memory pool}
Finally, we discuss a globally shared memory pool.
It is depicted in \cref{fig:mem-pool}.
We identify a gap between the CXL \spec and available hardware.
Currently, there is no CPU or pool device that implements CXL 3.0 back invalidation flows, so cache-coherent sharing is unavailable \cite{CXL-pool-pcie-pooling-2025}.
Therefore, we distinguish between a non-coherent, realistic memory pool and a cache-coherent pool envisioned according to the \spec.


\MODEL does not apply to the non-coherent setting, because the central assumption of cache coherence is not satisfied.
This means that concurrent program executions would violate \MODEL's global cache invariant.
Bypassing caches, i.e. only allowing the \MODEL primitives \texttt{MStore}, \texttt{LOAD-from-M}, and \texttt{M-RMW}, retains correctness.
However, this would likely result in a massive performance loss.
Compensating for the lack of hardware guarantees with software is an interesting open research question \cite{shared-mem-coherence-1989, gam-2018}, particularly in the face of potentially partially coherent pools.
We note that, while CXL0 assumes cache coherence, it is agnostic about whether this is implemented by hardware or software.
If a suitable software protocol exists with a similar load/store interface and coherence guarantees, then CXL0 applies to it as well.


Naturally, \MODEL applies to the fully cache-coherent version.
The memory pool is modeled as an extra memory node.
Per the CXL \spec, interactions with remote caches and remote RMWs are unavailable, so \texttt{RStore}, \texttt{LOAD-from-C}, \texttt{LFlush}, \texttt{Propagate-C-C}, and remote \texttt{RMW}s are excluded.

\paragraph{Future configurations}

We see that current CXL configurations are limited compared to the general operational semantics of \MODEL.
However, we have ensured that \MODEL captures the evolving nature of CXL configurations up to coherent memory sharing and fully symmetric capabilities between hosts and devices.
\remove{
We provide a mapping from CXL transactions to \MODEL primitives in \cref{sec:cxl0_mapping}.
Furthermore, the current CXL \spec does not support some transactions captured in \MODEL, such as \texttt{RStore} from the host-side.
However, the choice of \MODEL primitive makes a big difference for performance, as we show in \cref{sec:cxl0_latency}.
Therefore, we advocate to extend the \spec and expose precise ISA-level controls, as a larger set of primitives gives the programmer finer options for data placement and expands the design space for correctness and performance.
}


\section{Practical Implications}
\label{sec:implications}


Next, we conduct a set of experiments to evaluate the practical implications of the \MODEL model on real hardware.
The goal is to determine how concrete CXL transactions map to \MODEL primitives, and to measure the latency of individual \MODEL primitives in isolation.


Our setup consists of an x86 CPU (host) and an FPGA (device), both running CXL 1.1, with a protocol analyzer connected between them.
The CPU accesses FPGA memory via CXL.mem, and the FPGA accesses CPU memory via CXL.cache.
Therefore, both types of memory are shared and coherent. 
On the FPGA side, we use a proprietary CXL IP by Intel to configure it as a CXL Type 2 device.
Besides the IP's default behavior, we can add custom logic between the FPGA memory and the CXL IP.
This allows us to manipulate the CXL traffic by responding to requests or sending new ones according to our needs.
The analyzer is a \textit{Teledyne LeCroy's T516 Protocol Analyzer} for PCIe 5.0 and CXL, which supports all CXL sub-protocols and device types. 

This setup corresponds to the first variation depicted in \cref{fig:main}.
We generate and observe both CXL.mem and CXL.cache transactions, and cover all of \MODEL's primitives in the evaluation. 
As noted in \cref{sec:newmodel}, all other settings provide, at most, the same set of primitives.


\subsection{Mapping the CXL \spec to \MODEL}
\label{sec:cxl0_mapping}

\remove{First, we map CXL transactions to abstract \MODEL primitives.
The exact mapping is presented in \cref{table:cxl0_mapping}.}
\cref{table:cxl0_mapping} depicts the mapping from CXL transactions to abstract \MODEL primitives.
As host requests to cache device memory are handled by CXL.mem, and device requests to cache host memory are handled by CXL.cache, we analyze \MODEL from the perspectives of both the CPU and the FPGA.
The CPU is denoted as host and the the FPGA as device in the table. 
We distinguish between the types of memory to which the target address belongs:
\textit{Host-attached Memory (HM)} or \textit{Host-managed Device Memory (HDM)}.
We then create all possible pairs of cache coherence states for the host and the device, and all possible variations of available operations.
We use the CXL analyzer to observe the CXL transactions that are generated on the link.

We find that there is a many-to-one mapping from CXL.cache and CXL.mem transactions to the abstract primitives captured in \MODEL. 
For example, the first row of \cref{table:cxl0_mapping} represents a \MODEL \texttt{Read} primitive issued by the host CPU, e.g., a normal \texttt{load}.
We denote the cache coherence states of the host and the device before executing the \MODEL primitives as $(Host,Device)$ where $Host,Device \in \{ M, E, S, I \}$ and $*$ indicates any of the possible MESI states.
If it targets HM, then two possible CXL transactions can be triggered.
If $(Host,Device) \text{ is } (*,I)$ then no CXL transaction is observed (None).
Else, $(Host,Device)$ is one of $(S,S), (I,S), (I,E) ,(I,M)$, and we observe a CXL.cache Host-to-Device (H2D) SnpInv request.
\remove{\begin{enumerate*}
    \item If the given cache coherence states for the host and the device are $(*,I)$, then we observe no CXL transactions (None).
    \item If the given cache coherence state is of the remaining ones, namely $(S,S)$, $(I,S)$, $(I,E)$ or $(I,M)$, we observe a CXL.cache Host-to-Device (H2D) SnpInv request.
\end{enumerate*}}
Similarly, when a \MODEL \texttt{Read} primitive targets HDM, we observe two possible CXL transactions, again. 
If $(Host,Device) \text{ is } (I,*)$, a CXL.mem Master-to-Subordinate (M2S) MemRdData request occurs.
If $(Host,Device)$ is $(S,S)$, $(S,I)$, $(E,I)$ or $(M,I)$, no CXL transaction is observed (None).

In addition, not all \MODEL primitives are available (\texttt{???} in \cref{table:cxl0_mapping}).
For instance, neither an \texttt{RStore} nor an \texttt{LFlush} can be generated on the CPU by a single instruction or sequence of instructions.
Although the FPGA provides extra flexibility for generating CXL transactions, invoking an \texttt{LFlush} still 
remains impossible since the proprietary IP does not offer enough control over the underlying CXL link to issue new custom CXL transactions which are not defined by the \spec.
These limitations restrict the programmer's control over the underlying CXL link, rendering those primitives currently unavailable for algorithmic use.
This is a critical shortcoming because the choice of \MODEL primitive makes a big difference to performance, as we will show next.


\begin{table*}
    \centering
    \small
    \begin{tabulary}{\textwidth}{|C|c|C|C|C|C|C|}
    \hline
    \rowcolor[HTML]{B3B3B3}
        \textbf{CXL0 primitive} & \textbf{Node} & \textbf{Operation} & \textbf{to HM} & \textbf{to HDM in Host-Bias}\\
    \rowcolor[HTML]{E6E6E6}
    \hline
         & & \textbf{x86 Instructions} & \textbf{CXL.cache H2D} & \textbf{CXL.mem M2S} \\
    \hline
        \texttt{Read} & \multirow{6}{*}{Host} & Load & None, SnpInv & None, MemRdData \\\cline{1-1}\cline{3-5}
        \texttt{LStore} & & Store & None, SnpInv & None, MemRdData, MemRd \\\cline{1-1}\cline{3-5}
        \texttt{RStore} & & \texttt{???} & \texttt{???} & \texttt{???} \\\cline{1-1}\cline{3-5}
        \texttt{MStore} & & Non-Temporal Store + Fence & SnpInv & MemWr \\\cline{1-1}\cline{3-5}
        \texttt{LFlush} & & \texttt{???} & \texttt{???} & \texttt{???} \\\cline{1-1}\cline{3-5}
        \texttt{RFlush} & & CLFlush & None, SnpInv & None, MemInv, MemWr \\
    \hline
    \rowcolor[HTML]{E6E6E6}
         & & \textbf{Device Operation} & \textbf{CXL.cache D2H} & \textbf{CXL.cache \& CXL.mem} \\
    \hline
        \texttt{Read} & \multirow{6}{*}{Device} & Caching Read & None, RdShared & None, RdShared \\\cline{1-1}\cline{3-5}
        \texttt{LStore} & & Caching Write & None, RdOwn & None, RdOwn \\\cline{1-1}\cline{3-5}
        \texttt{RStore} & & HM: ItoMWr / HDM: Caching Write & ItoMWr & None, RdOwn \\\cline{1-1}\cline{3-5}
        \texttt{MStore} & & Caching Write + CLFlush & (RdOwn +) DirtyEvict, WOWrInv/F, WrInv & None, MemRd \\\cline{1-1}\cline{3-5}
        \texttt{LFlush} & & \texttt{???} & \texttt{???} & \texttt{???} \\\cline{1-1}\cline{3-5}
        \texttt{RFlush} & & CLFlush & CleanEvict, DirtyEvict & None, MemRd \\
    \hline
    \end{tabulary}
    \vspace{8pt}
    \caption[Observable CXL requests for all possible \MODEL primitives.]{
        Observable CXL transactions for all possible \MODEL primitives. \remove{from both the CPU \textit{(top)} and the Device \textit{(bottom)} going to both Host-attached Memory (HM) and Host-managed Device Memory (HDM) in host-bias for all cache coherence states.}
    }
    \label{table:cxl0_mapping}
\end{table*}

\subsection{Latency Measurements} 
\label{sec:cxl0_latency}

To measure the latency of individual \MODEL primitives in isolation, we make the following configuration:
\begin{itemize}[leftmargin=*,itemsep=0pt]
    \item For \texttt{Load} instructions, the initial state of a cache line is \textsc{invalid} in all caches.
    \item For \texttt{Store} instructions, full cache line writes are performed.
    \item To measure latency from the CPU side, we use LATTester~\cite{LatTester} 
    \item To measure latency from the FPGA side, we implement custom logic.
          To execute a new memory load/store request using CXL, we issue an AXI-bus request to the CXL IP.
          Upon completion, the CXL IP responds via the designated read/write-response channels of the AXI-bus.
          Counting the number of FPGA clock cycles that elapse between the two events allows us to determine the latency of a \MODEL primitive on the device.
\end{itemize}


\begin{figure}[t]
    \vspace{-5pt}
    \centering
    \includegraphics[width=1.05\columnwidth]{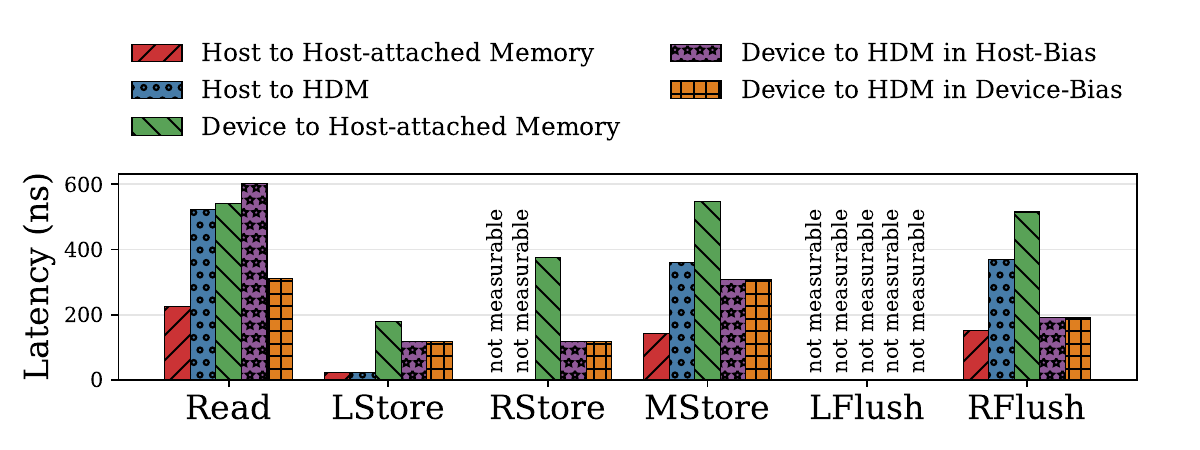}
    \caption{Latency of \MODEL primitives on host and device.}
    \label{fig:experiment_latency}
    \vspace{-10pt}
\end{figure}

\Cref{fig:experiment_latency} depicts latency measurements for various \MODEL primitives.
We characterize five types of accesses.
Host accesses to (1) Host-attached Memory (local) or (2) Host-managed Device Memory (HDM, remote), and device accesses to (3) Host-attached Memory (remote), (4) HDM in host-bias (local, but requires permission from the host) or (5) HDM in device-bias (local).
We report the median over 1000 measurements of sequential memory accesses.

We observe that loads and stores to local memory are twice as fast as those to remote memory for \texttt{Read} and \texttt{MStore} primitives.
For the CPU, it is $2.34$x faster, and for the device, it is $1.94$x faster (see \textit{red} vs. \textit{blue} and \textit{orange} vs. \textit{green} bars).
These results are consistent with previous findings \cite{sharma2023introduction, againstCXL}.
Furthermore, accesses from the host and the device to their respective remote CXL memory yield the same latency, despite using different CXL sub-protocols.

We observe our predicted latency trends for store operations.
\texttt{LStore} is unsurprisingly fast.
However, there is a difference between the \texttt{LStore} writes on the CPU and FPGA (\textit{red} and \textit{blue} vs. other \texttt{LStore}s).
The CPU takes advantage of its write buffers, while the device employs a single level of caches and no specialized hardware.
\texttt{LStore} primitives issued by the FPGA also differ, with slower local cache writes to Host-attached Memory (\textit{green}) than to HDM (\textit{purple} and \textit{orange}).
This is due to the CXL IP using two separate caches of different size, depending on the memory target.
When a device writes to Host-attached Memory (\textit{green} bars), \texttt{MStore} is $1.45$x slower than \texttt{RStore}, which is, in turn, $2.08$x slower than \texttt{LStore}.
These differences are expected, as access to more distant memory incurs higher latency, and this trend should persist in subsequent CXL versions.
\texttt{RFlush} is the only type of flush that both the CPU and the FPGA can invoke.
The measured latencies are nearly identical to those of the \texttt{MStore} primitive.

To summarize, we provide a mapping from CXL transactions to \MODEL primitives, and find that the CXL \spec currently does \emph{not} support all transactions captured in \MODEL, such as \texttt{RStore} from the host-side or \texttt{LFlush} from either side.
However, we also show that the choice of \MODEL primitive \emph{significantly} affects performance.
Therefore, we advocate to extend the \spec and expose precise ISA-level controls, similar to flushes in x86, as a larger set of primitives gives the programmer finer options for data placement and expands the design space for correctness and performance.

\section{General Transformations}\label{sec:transformation}

Ensuring correct execution of concurrent programs under spontaneous failures is a challenging task~\cite{chandi1986dist, elnozahy2002dssurvey, koo1987checkdist, stevens16checkpoint}.
To alleviate the programmer's burden, we present a transformation that provides concurrent algorithms with fault tolerance guarantees under \MODEL.
Our transformation is an adaptation of FliT by \citet{wei2022flit}.
We show that with minor adjustments, FliT can be applied to a broader range of use cases than originally intended.
This transformation equips \emph{any} linearizable object with durable linearizability under \MODEL.
First, we present a simple example that clearly illustrates the need for a new set of transformations.
Then, we define the desired correctness guarantees in the \MODEL model.
Next, we provide an introduction to FliT and discuss our adaptation of it.
Throughout this section, we assume that all shared memory is either non-volatile or in a different failure domain.
In addition, we do not assume a specific memory model.
All of the presented algorithms list instructions in \emph{program order}, so additional architecture-specific memory fences may be necessary (e.g., \texttt{MFENCE} for x86, and \texttt{DMB} for ARMv8) when implementing them on real hardware.




\paragraph{Motivating example}

To emphasize the difference between the single-machine setting modeled in \cite{izraelevitz2016linearizability} and the distributed setting with CXL, we present a simple litmus test.

In the program below, $r1$ and $r2$ are local registers and $x$ is a shared variable initialized to 0.
The program runs on machine M1 and $x$ is allocated on machine M2 ($x \in Loc_{M2}$).
\begin{align}
&\texttt{x=1;}\hspace{16pt}\texttt{r1=x;}\hspace{16pt}\texttt{r2=x;}\hspace{16pt}\texttt{assert(r1==r2);} &\text{\myxmark}
\end{align}
We expect $r1$ and $r2$ to be equal, but our system model allows for different behaviors, e.g., if M2 crashes and recovers between the execution of the two load steps.
When M1 updates $x$, the update resides in its cache, then propagates to M2’s cache, and finally to M2’s memory, where it persists.
If M2 crashes after the update has left M1’s cache but before it reaches M2’s memory, the update is lost.
Thus, a remote machine’s crash can affect the correctness of a local program, something that doesn’t occur in single-machine settings.

The same would hold true if a flush which evicts $x$ from M1's cache is placed between the first two commands, or if an additional flush is placed between the reads of $x$.
In the full-system crash model, this inconsistency is impossible, and either of the aforementioned flushes would preserve the persistence ordering of future reads with respect to the update of $x$. 
However, under \MODEL, we require a flush which ensures that the data reaches physical memory in order to guarantee that the assertion is always satisfied.

To conclude, the partial-system crash model introduces consistency challenges, and there is a need to distinguish between flush operations with different post-conditions.

\paragraph{Correctness Guarantees}

Our transformation provides \emph{durable linearizability}, a correctness criterion presented by \citet{izraelevitz2016linearizability} in a full-system crash model.
We discuss the applicability of the definition to our partial-system crash model, review the key steps in the definition's formulation and show that it applies to \MODEL's system model as is.

The abstract model by \citet{izraelevitz2016linearizability} treats high-level objects and considers the invocation and response of operations on these objects as events in the system's history.
In our model, we consider partial-system crashes, e.g., single-machine failures, as events as well. 
Machines may accommodate multiple running threads that crash together and instantaneously.
As in \cite{izraelevitz2016linearizability}, we assume that, when recovering from a crash, new threads with new and distinct identifiers are spawned to replace the ones that operated prior to the crash.
However, we limit this to the crashed machine, i.e., threads running on other machines are uninterrupted.
Simultaneous failures of any combination of machines is captured in \MODEL by appending local crashes one after the other, in any order.


We define an \emph{abstract history} as a sequence of invocation, response, and single machine crash events.
All the instructions that compose into a high-level operation take place between its invocation and response.
We keep the definition of abstract well formedness as originally defined in \cite{izraelevitz2016linearizability}.
A \emph{well formed history} is a history in which the local history of every thread is a properly interleaved sequence of invocation and response events, possibly ending with a local crash. 


A history is \emph{durably linearizable} if it is well formed, and linearizable after all crash events are removed from it.
Linearizability itself is defined using the \emph{abstract happens-before relation} between events, as described in~\cite{herlihy1990linearizability,Sela21}, which is the partial order obtained by restricting the real-time order in a history to consist of the order between events of the same thread, as well as the order from a response event to an invocation event.
We remind that the definition of linearizability allows both completing or omitting pending invocations without matching responses.

While \citet{izraelevitz2016linearizability} redefine the ``abstract happens-before relation'' so it orders events also with respect to crashes, we observe that this is unnecessary, as the history checked for linearizability is a crash-free history obtained from the original history by removing all the crash events. Thus, there is no need to modify the relation, and we keep the original definition of~\cite{herlihy1990linearizability,Sela21} as is.
Therefore, the definition of durable linearizability is agnostic about the scope of a crash and applies to our model without modification.
From this point on, we use the term ``durably linearizable'' in the context of \MODEL's system model and failure model, which captures crashes at single machine granularity.

We recall that linearizability and durable linearizability are local, composable properties.
In other words, combining (durably) linearizable objects yields (durably) linearizable histories.
This locality has immense practical implications, as objects designed to satisfy some local property can be used together without further reasoning or adaptations, and the resulting executions will also satisfy that property.

\paragraph{Introduction to FliT}
We seek to devise a general and correct transformation that satisfies durable linearizability and chose to adapt FliT by \citet{wei2022flit}.
FliT takes linearizable objects and makes them durably linearizable in the full-system crash model.
Among existing transformations that extend objects with durable linearizability in the single-machine, full-system crash model (\cref{sec:related}), FliT is general as it only requires the original object to be linearizable.
Following prior FliT proofs \cite{wei2022flit, bodenmuller2023fully}, we separate liveness from safety.
Our work focuses exclusively on durable linearizability, which is a safety property, and leaves liveness to the programmer.
Therefore, just like original FliT, we assume the original data structure to be non-blocking to trivially guarantee liveness.

Due to the locality of durable linearizability, it is possible to transform every object independently to ensure that an algorithm spanning multiple objects inherits its correctness from the underlying objects. 
It is important to note that FliT transforms objects that are \emph{already} linearizable, i.e., synchronization over high-level objects is handled at the object level.
Consequently, FliT does not introduce extra concurrency control, nor does our adaptation.

At the core of FliT is a classification of memory accesses.
FliT distinguishes between shared and private memory operations.
Private memory operations operate on data that will never be accessed by more than one thread concurrently.
Examples include thread-local counters and logs.
Conversely, shared operations span shared data that may be accessed concurrently by multiple threads.
Furthermore, FliT also differentiates between operations that need to become persistent and those that do not.
To avoid blindly persisting every memory write, FliT uses a persistence flag (\texttt{pflag}) to tag only the necessary operations.
This flag indicates whether the address is in persistent memory, and whether updates to that address should be durably linearizable.
Only tagged operations are handled with the proper flush instructions.

The original version of the FliT transformation assumes execution on an x86 processor and is depicted in \cref{alg:Flit_org}.
With x86, FliT assumes the total-store-order memory model (TSO), and that a \texttt{Flush} corresponds to an optimized asynchronous flush instruction (e.g., \texttt{CLFLUSHOPT}).
FliT uses \texttt{SFENCE}, which acts as a memory barrier that no store operation can cross in any direction.
Prior flush operations must be completed before a thread can perform store instructions after this barrier.

\begin{figure}[t]
\centering
\begin{lstlisting}[language=C++,caption=The Original FliT Transformation (for x86), label=alg:Flit_org]
T private_load(T* X) { 
	return X->Load();
 }
	
void private_store(T* X, T args, bool pflag) {
	if (pflag) {
		X->Store(args);(*@\label{line:flit:ppstore}@*)
		Flush(X); 
		MFENCE(); 
	} 
   else X->Store(args);(*@\label{line:flit:pvstore}@*)
 }

T shared_load(T* X, bool pflag) {
	T val = X->Load();
	if (pflag && flit_counter(X) > 0) Flush(X); (*@\label{line:flit:sload}@*)
	return val;
 }
 
void shared_store(T* X, T args, bool pflag) {
	FENCE(); (*@\label{line:flit:sstorefence1}@*)
	if (pflag) {
		flit_counter(X).fetch&add(1);(*@\label{line:flit:inc}@*)
		X->Store(args);
		Flush(X); (*@\label{line:flit:sstoreflush}@*)
		MFENCE();(*@\label{line:flit:sstorefence2}@*)
		flit_counter(X).fetch&sub(1);(*@\label{line:flit:dec}@*)
	}
	else X->Store(args); 
 }
 
void completeOp() { 
	MFENCE(); (*@\label{line:flit:completeop}@*)
 }
\end{lstlisting}
\vspace{-15pt}
\end{figure}

The \texttt{private_load}, \texttt{private_store}, \texttt{shared_load}, and \texttt{shared_store} methods serve as wrappers for memory accesses that provide persistence and durable linearizability with the help of \texttt{completeOp}.
The \texttt{private_load} method performs a load of the specified object.
If flagged as persistent, a \texttt{private_store} of an object stores it in persistent memory by performing a \texttt{Store} operation (\cref{line:flit:ppstore}) followed by a \texttt{Flush} to write the value to memory.
Otherwise, it performs a \texttt{Store} (\cref{line:flit:pvstore}), which has no guarantees with respect to persistence, so its propagation to memory may be reordered with those of later stores.  
The \texttt{shared_load} and \texttt{shared_store} methods satisfy a guarantee stronger than just persistence of flagged operations: durable linearizability.
Because a store operation is not immediately persistent after execution, it may become globally visible without being guaranteed to survive a crash.
FliT ensures that, when a value is read, it is already persistent by the time the high-level operation containing that read completes and before any subsequent \texttt{shared_store}.

To avoid na\"ively flushing every location upon reading, FliT uses a shared counter for each shared object, called the \emph{FliT counter}. 
This counter signals to readers that a store is being performed, but it is not necessarily persistent yet.
Shared store operations increment the counter (\cref{line:flit:inc}) before applying the store to shared memory and persisting it (\cref{line:flit:sstoreflush,line:flit:sstorefence2}), and decrement it after the store persists (i.e., after a corresponding full memory barrier, \cref{line:flit:dec}). 
A shared load operation checks if the counter's value is positive.
If so, it helps to propagate the stored value by flushing it without issuing a \texttt{FENCE} instruction.
The complementing \texttt{FENCE} instructions are located in the \texttt{completeOp} method and at the beginning of a \texttt{shared_store} (\cref{line:flit:sstorefence1}). 
FliT's methods take an additional argument (\texttt{pflag}) to distinguish between tagged and non-tagged operations.
This flag indicates whether the address is in persistent memory, and whether updates to that address should be durably linearizable.
If \texttt{pflag} is not set, operations do not flush the object to persistent memory.

Lastly, FliT facilitates \texttt{completeOp}, a method which guarantees that all effects of the high-level operation are persisted before the operation returns, as asserted by the definition of durable linearizability.
This method consists of a single memory fence (\cref{line:flit:completeop}) and guarantees that flushes executed by \texttt{shared_load} operations take effect before it returns. 
Write persistence is already guaranteed at the end of the store operation itself.

When using FliT, shared and private memory access methods are expected to wrap concrete memory accesses of the high-level object, and a \texttt{completeOp} is expected to be placed at the end of every high-level object operation.

\subsection{Adapting FliT to \MODEL}

As shown in the brief example, FliT cannot be adopted as is in \MODEL, due to the semantics of the store and flush instructions that it uses.
In addition, FliT employs asynchronous flushes, deferring the time where reads are \emph{guaranteed} to be flushed until the following memory fence.
Hence, it is vulnerable and must be adapted for partial-system crashes.

Our transformation is designed for high-level code.
We assume support for \MODEL's load, store, and flush primitives in the form of respective high-level instructions or methods.
We also assume that there exists an atomic fetch-and-add (\texttt{FAA}) operation, and remind that \MODEL captures the behavior of \texttt{RMW} operations as part of its operational semantics.
Unlike the original FliT, we do not assume a fixed memory model.

\begin{figure}[t]
\centering
\begin{lstlisting}[language=C++, caption=FliT Transformation for \MODEL, firstnumber=last, label=alg:flit-adapt]
T private_load(T* X) {
    return X->Load();(*@\label{line:pload}@*)
}
	
void private_store(T* X, T args, bool pflag) {
	if (pflag) {
		(*@\colorbox{backcolour}{X->LStore(args);}@*) (*@\label{line:pstore}@*)(*@\label{line:replace_LR_1}@*)
		(*@\colorbox{backcolour}{RFlush(X);}\label{line:rm4}@*)(*@\label{line:prflush}@*)
	} 
   else (*@\colorbox{backcolour}{X->LStore(args);}@*) 
}

T shared_load(T* X, bool pflag) {
	T val = X->Load();(*@\label{line:sload}@*)
	if (pflag && flit_counter(X) > 0) (*@\colorbox{backcolour}{RFlush(X);}\label{line:rm1}@*)
	return val; 
}

void shared_store(T* X, T args, bool pflag) {
	if (pflag) {
		flit_counter(X).fetch&add(1);
		(*@\colorbox{backcolour}{X->LStore(args);}@*) (*@\label{line:replace_LR_2}@*)(*@\label{line:sstore}@*)
		(*@\colorbox{backcolour}{RFlush(X);}@*) (*@\label{line:rm3}@*)(*@\label{line:srflush}@*)
		flit_counter(X).fetch&sub(1);
	}
    else (*@\colorbox{backcolour}{X->LStore(args);}@*) 
}

void completeOp() { }
\end{lstlisting}
\end{figure}



\cref{alg:flit-adapt} depicts our adaptation of FliT.
All \texttt{Flush} instructions are replaced with \texttt{RFlush}, and all stores, whether marked for persistence or not, use \texttt{LStore} instead of the architecture's store instruction.
The choice of \texttt{RFlush} ensures that locally cached updates become persistent before the high-level operation is completed, thereby achieving durable linearizability.
We mark the translated lines with a colored background.
In \cref{sec:flit-proof} of the supplementary material, we prove that our transformation yields durably linearizable objects when applied to linearizable ones.



The \texttt{completeOp} method is empty in our transformation. 
In the original FliT, we recall that the method comprises a single memory fence.
This fence is unnecessary when assuming both in-order execution and synchronous flushes.
However, when applying our transformation to real hardware with weaker memory models, \cref{alg:flit-adapt} should employ appropriate fences to enforce correct ordering.

\paragraph{Implementation and performance} 
\texttt{LStore} and \texttt{RFlush} are available in all cache-coherent settings of the system model evolution, which highlights the applicability of this transformation to a wide variety of current and future CXL setups.
Generally, it is possible to substitute one sequence of primitives for another that is semantically equivalent (see \cref{prop:one}).
For instance, replacing all stores with \texttt{MStore} automatically satisfies the desired durability guarantees, even in systems without cache coherence.
However, this is expected to yield inferior performance, and even the CXL \spec advises using weaker transactions for better performance (\S 3.5.2.3 in \cite{CXLSpec40}).
One way to achieve this is by exploiting additional address information.
For example, \texttt{RFlush} can be replaced with \texttt{LFlush} for all memory locations owned by the writing machine.
In addition, on hardware with implicit write-back on remote loads, FliT counters can be kept in local memory.
Unlike the original FliT, our adaptation allows automatic instrumentation of the transformation based solely on the memory address.
However, as with FliT, manual tuning is still required to reduce the amount of flush instructions for optimal performance.
\section{Related Work}\label{sec:related}

A variety of work has provided formal semantics for shared-memory models that were originally specified in prose or internal engineering discussions.
This includes seminal work for x86~ \cite{SewellSONM10,OwensSS09}, other multi-core architectures~\cite{AlglaveMT14,FlurGPSSMDS16}, GPU concurrency~\cite{LustigSG19,lc-memento}, non-volatile memory with full-system crashes~\cite{RaadWNV20}, and remote memory access (RMA)~\cite{DanLHV16}.
Our work falls within this scope.
However, given that CXL~3.0 hardware is not currently available, we cannot test our model against actual implementations.
Thus, our evaluation is confined to version 1.1 of the CXL standard (see \cref{sec:implications}).

\citet{goens2023compound} describe how specific memory models fuse together, focusing on PTX and x86. 
Our work proposes a higher-level programming model on top of CXL and is agnostic about the memory model's internals.

Multiple attempts have been made to provide concurrent algorithms with durability guarantees. 
Our work is inspired by~\citet{izraelevitz2016linearizability}, who propose a programming model for persistent memory, study correctness notions under this model, and suggest the first general construction for linearizable objects.
However, their construction has limited applicability because of its performance overhead.
Since then, several more specialized constructions have been proposed.
NVTraverse~\cite{nvtraverse} targets lock-free traversal data structures.
Mirror~\cite{mirror} is an automatic transformation for any lock-free data structure.
FliT~\cite{wei2022flit} transforms any linearizable object without restriction, as discussed before.
Montage~\cite{DBLP:conf/icpp/Wen0DJVS21} guarantees recoverability to \emph{some} consistent state, but not necessarily to the most recent one prior to the crash.
These constructions are designed for the full-system crash model and local persistent memory.
In this paper we show how to transform such objects to work in the more general partial-system failure model and with disaggregated memory.


Some systems are CXL ready.
However, they neither provide a general programming model nor a transformation for existing algorithms.
Āpta~\cite{apta} is a design for function-as-a-service (FaaS) over CXL, it does not target shared memory and does not consider multiple architectures.
\citet{pond} designed Pond, a memory pooling system for CXL.
\citet{10.1145/3582016.3582063} propose TPP for OS-level page placement on top of CXL.
Lupin \cite{lupin-2024} helps distributed applications tolerate partial failures using a shared CXL pod for replication.
These  system designs are orthogonal to this work. 

Frameworks and systems for general disaggregated memory architectures enjoy increasing interest in recent years.
\citet{mind} propose an RDMA based memory management system.
Teleport~\cite{zhang2022optimizing} and Cowbird~\cite{cowbird} offer techniques for better distributing work across the system.
\citet{10.1145/3445814.3446713} develop a software runtime that aims to improve memory access times.
However, these do not target CXL and aim to improve either work distribution or memory access patterns.
By contrast, our work contributes a correct general transformation for any concurrent algorithm.

\section{Conclusion and Future Work}
\label{sec:conc}
This work introduces \MODEL, the first programming model designed for disaggregated memory using CXL technology.
We demonstrate the applicability of \MODEL to current and future CXL setups, and illustrate its usefulness by developing a transformation that enables linearizable and durably linearizable concurrent algorithms to function correctly within a CXL environment.
Adapting existing algorithms to the new model allows both researchers and practitioners to explore new opportunities in this design space. 

We consider \MODEL to be a solid foundation that can be refined in various ways. We have already discussed two variants in \cref{sec:variants}.
It would be useful to extend \MODEL with  heterogeneous memory settings with non-CXL, privatized memory to capute more application use-cases.
Another interesting refinement would be to distinguish between synchronous (posted, in PCIe terminology) and asynchronous (non-posted) CXL transactions (see \S 5.2 of the survey by \citet{sharma2023introduction}).
Finally, relaxing durability semantics has generally been shown to be beneficial for performance \cite{assa2023tl4x,DBLP:conf/icpp/Wen0DJVS21} and can be explored here as well.
Enhancing \MODEL with these perspectives will result in a richer model and potentially better-performing transformations and algorithms.
\newpage
\bibliographystyle{ACM-Reference-Format}
\bibliography{references}

\clearpage
\setcounter{page}{1}
\appendix

The supplementary material is organized as follows.
\cref{sec:prop_proof} provides proofs for items 7 and 8 of \cref{prop:one}, and \cref{sec:flit-proof} has the correctness proofs for our adaptation depicted in \cref{sec:transformation}.

\section{Proofs of Items 7 and 8, Proposition 1}
\label{sec:prop_proof}

For $(7)$, we first observe that it suffices to show that
{\scriptsize $\conf_1~\xrightarrow{\texttt{LStore}_j(x,v)}~\conf_2$}, 
{\scriptsize $\conf_2~\xrightarrow{\tau^*}~\conf_3$}, 
and {\scriptsize $\conf_3~\xrightarrow{\texttt{LFlush}_j(x)}~\conf_4$}
together imply that
\smallskip
{\scriptsize $\conf_1~\xrightarrow{\texttt{RStore}_j(x,v)}~\conf_4$}.
The proof of this claim proceeds by induction on the length of the trace from $\conf_2$ to $\conf_3$.
If this trace starts with a propagation (either \texttt{C-C} or \texttt{C-M}) of a location $y$ different from $x$, we can commute this propagation before the \texttt{LStore}, and derive
{\scriptsize $\conf_1~\xrightarrow{\texttt{RStore}_j(x,v)}~\conf_4$}
from the induction hypothesis.
If this trace starts with a propagation of $x$, it must be a \texttt{C-C} propagation in node $i$.
Then, the effect of $\texttt{LStore}$ followed by this propagation is similar to the effect of $\texttt{RStore}$.
Finally, we observe that
{\scriptsize $\conf_3~\xrightarrow{\texttt{LFlush}_j(x)}~\conf_4$}
ensures that the analyzed trace must contain a propagation step of $x$.

Item $(8)$ is proved similarly, but requires another nested induction. 
Again, we observe that it suffices to show that
{\scriptsize $\conf_1~\xrightarrow{\texttt{LStore}_j(x,v)}~\conf_2$}, 
{\scriptsize $\conf_2~\xrightarrow{\tau^*}~\conf_3$},
and {\scriptsize $\conf_3~\xrightarrow{\texttt{RFlush}_j(x)}~\conf_4$}
together imply that
\smallskip
{\scriptsize $\conf_1~\xrightarrow{\texttt{MStore}_j(x,v)}~\conf_4$}.
We prove this claim by an induction similar to the one above.
For the case that the middle trace of propagations starts with a \texttt{C-C} propagation of $x$, we observe that the effect of this propagation after the $\texttt{LStore}$ is identical to an $\texttt{RStore}$, and we use another similar induction argument to show that
{\scriptsize $\conf_1~\xrightarrow{\texttt{RStore}_j(x,v)}~\conf_2$}, 
{\scriptsize $\conf_2~\xrightarrow{\tau^*}~\conf_3$},
and {\scriptsize $\conf_3~\xrightarrow{\texttt{RFlush}_j(x)}~\conf_4$}
together imply that
{\scriptsize $\conf_1~\xrightarrow{\texttt{MStore}_j(x,v)}~\conf_4$}.

\section{Correctness Proofs}
\label{sec:flit-proof}

In this section we prove the correctness of our adapted FliT transformation under \MODEL.
\citet{wei2022flit} define the \emph{P-V Interface} as listed in \cref{def:spec}, and use it to prove the correctness of the original FliT transformation.
We follow the same methodology in order to prove the correctness of our adaptation.
The P-V interface captures and formalizes the dependencies between threads and persistent store operations.
It distinguishes between p-instructions and v-instructions.
The former have to be persisted, while the latter do not.
Both types can be load or store operations denoted as \emph{p-load}, \emph{v-load}, \emph{p-store}, and \emph{v-store}.

\begin{definition}[The P-V Interface~\cite{wei2022flit}]
\label{def:spec}
	Each instruction has a linearization point within its interval, such that:
	\begin{enumerate}[leftmargin=*,itemsep=0pt]
		\item \label{cond:seqSpec} A load $r$ on location $\ell$ returns the value of the most recent store on $\ell$ that linearized before $r$.
		\item \label{cond:persistStore} Let $s$ be a linearized \emph{p-store} executed by a thread $i$. $i$ depends on $s$. 
		\item \label{cond:persistLoad} Let $r$ be a \emph{p-load} by thread $i$ on location $\ell$. $i$ depends on every \emph{p-store} on $\ell$ that was linearized before $r$.		
		\item \label{cond:fence} Let $t$ be either the linearization point of a shared store by thread $i$, or the time at which $i$ completes an operation. The value of every store that $i$ depended on before time $t$ is persisted by time~$t$. 
	\end{enumerate}
\end{definition}

The \emph{interval} of an operation is the time between its invocation and completion.
Intuitively, we say that a thread $i$ \emph{depends on} a store operation \emph{op} if:
(1) \emph{op} is a store performed by $i$, or
(2) \emph{op} is a store that happened before a load to the same location by $i$.

We now demonstrate that \cref{alg:flit-adapt} satisfies the P-V interface.
According to Theorem 3.1 in~\cite{wei2022flit}, if every load and store are p-instructions, then an algorithm that satisfies the P-V interface is durably linearizable.
This result has also been formally verified by \citet{bodenmuller2023fully}.
The aforementioned theorem remains correct in \MODEL's failure model, since
(1) the definition of durable linearizability applies to \MODEL as is and does not use the crash event in its formulation \remove{(see also \cref{remark:one})}, and
(2) neither the theorem nor the P-V interface considers where data resides or which parts of the system crash.
They only assert that dependencies persist in the correct order.

Note that it is the value of \texttt{pflag}, and not the type of store instruction in use, that determines whether a write or read is a p-instruction or not.
We show that if \texttt{pflag} is always set, any algorithm transformed by \cref{alg:flit-adapt} satisfies all the conditions specified in \cref{def:spec}.
Our approach is similar to that of \citet{wei2022flit} in the correctness proof of the original FliT transformation.

\begin{proof}
    We set the linearization point of any memory operation that operates on address $x$ to be the time at which the actual load or store (that is, \cref{line:pload,line:pstore,line:sload,line:sstore} \remove{, line:sstore}) is executed. To satisfy Condition~\ref{cond:fence} and persist dependencies on time, we show that an \texttt{RFlush} is executed for each dependency defined in Conditions~\ref{cond:persistStore} and~\ref{cond:persistLoad}, and that it takes effect before the time dictated by Condition~\ref{cond:fence}. We prove that \cref{alg:flit-adapt} satisfies Conditions \ref{cond:seqSpec} and \ref{cond:fence} and respects the dependencies of Conditions \ref{cond:persistStore} and \ref{cond:persistLoad} in the order in which they appear in \cref{def:spec}.

	\textsc{Condition~\ref{cond:seqSpec}.} Each transformed load or store operation, whether private or shared, executes exactly one low-level load or store instruction, respectively. These memory accesses happen under the same synchronization guarantees of the high-level object, with the addition of the FliT counter. Thus, we do not introduce new behaviors that were not permitted under the original linearizable implementation. This, combined with the cache coherence property, satisfies Condition \ref{cond:seqSpec}.

	\textsc{Condition~\ref{cond:persistStore}.} The \emph{p-store}s of a high-level operation are persisted before the operation completes. Each \emph{p-store} is followed by a synchronous flush operation (\cref{line:prflush,line:srflush}), after which it is guaranteed to reside in persistent memory. Due to in-order execution, further progress after the flush is impossible, and hence the \emph{p-store} is guaranteed to be persistent before any other \emph{p-store} by the same high-level operation and before the operation completes. 

	\textsc{Condition~\ref{cond:persistLoad}.} Consider a \emph{p-load} of address $x$ executed by thread $i$. The load instruction executes regardless of whether the operation is private or shared. If it is a \texttt{private\_load}, there is no data race over address $x$. Thus, there is no need to propagate the value since the last \texttt{private\_store} did so before returning. These operations cannot execute concurrently by the definition of private operations.
 
    For \texttt{shared\_load}, however, we need to consider the FliT counter. It may be zero or a positive value. It cannot be negative since threads decrement the counter only after incrementing it and these operations cannot be reordered. If the value is zero, then the \texttt{shared\_store} that decremented the counter to zero has already persisted the stored value. In particular, this store linearized before the decrement took effect because of in-order execution. 
    
    If the value is positive, an \texttt{RFlush} is executed for address $x$. Note that the value which is flushed by \texttt{shared\_load} may be from a later store to $x$ than the one observed by the load. However, the combination of propagation and in-order execution ensures that any store to $x$, that linearized before the low-level load took place, is persisted before the \texttt{shared_load} returns. In particular, this occurs before the high-level operation completes.

    \textsc{Condition~\ref{cond:fence}.} Due to in-order execution, each \texttt{shared_store} operation takes effect before any other instruction by the same thread is executed. This means that any \texttt{RFlush} performed by the process as part of a \texttt{shared_load} has taken its effect before any dependent shared store executes. Also, a shared store persists only after the propagation of previous \emph{p-store}s by the same process. So, the \emph{p-store} persists only after all its dependencies. 

    \qedhere

\end{proof}

\paragraph{Note on weaker memory models and multiple machines}
The correctness of this proof holds for weaker memory models almost as is.
Assuming ordering primitives (i.e., fences) are placed properly, replacing the arguments which rely on in-order execution is trivial.
Moreover, since the entire proof is confined to arguments that regard local ordering, the transformation works even when executed by multiple machines with different architectures that share memory under the same assumption.

\end{document}